\documentclass[12pt]{article}
\usepackage{amssymb}
\usepackage{amsmath}
\usepackage{graphicx}
\usepackage{calrsfs}
\usepackage{cancel}
\usepackage{hyperref}
\usepackage{longtable}
\usepackage{MnSymbol} 

\newcommand{\be}{\begin{equation}}
	\newcommand{\ee}{\end{equation}}
\def\bea{\begin{eqnarray}}
	\def\eea{\end{eqnarray}}

\newcommand{\bn}{\begin{eqnarray}}
	\newcommand{\en}{\end{eqnarray}}

\newcommand{\p}{\partial}

\newcommand{\nn}{\nonumber}

\newcommand{\no}{\noindent}

\newcommand{\s}{\,\,\,\,}
\def \produtoria {\prod\limits}

\def \dtres {$D=3+1$\;}

\def \dtreS {$D=3+1$}
\def \ddoiS {$D=2+1$}

\def \produtoria {\prod\limits}

\def\bea{\begin{eqnarray}}
	\def\eea{\end{eqnarray}}
\def\beq{\begin{eqnarray}}
	\def\eeq{\end{eqnarray}}

\def \del{\partial}

\def\bea{\begin{eqnarray}}
	\def\eea{\end{eqnarray}}

\usepackage{xcolor}

\usepackage[normalem]{ulem}

\title{\textbf{Generators of the Poincaré Group for arbitrary tensors and spinor-tensors} }

\begin{document}
	\author{H.V. Almeida Silva$^a$ \footnote{hudson.vinicius@unesp.br}, D. Dalmazi$^a$ \footnote{denis.dalmazi@unesp.br}, R.R. Lino dos Santos$^b$ \footnote{rafael.santos@ncbj.gov.pl}, E.L. Mendonça$^a$ \footnote{elias.leite@unesp.br}\\
		\textit{{a- UNESP - Campus de Guaratinguetá - DFI }}\\
		\textit{{b- National Centre for Nuclear Research, Pasteura 7, 02-093 Warsaw, Poland}}\\
	} 
	\date{\today}
	\maketitle
	
	\begin{abstract}
		In this work, we systematically derive explicit expressions for the Poincaré Group generators on arbitrary-rank tensors and spinor-tensors in $D=3+1$ and $D=2+1$ spacetimes, thus generalizing previous works in the literature for the groups $ISO(3,1)$ and $ISO(2,1)$. From the Casimir eigenvalue equations, we demonstrate in a model-independent way the Fierz-Pauli constraints for massive particles for spins $\mathfrak{s}=5/2$, 3, and 4 in $D=3+1$ and helicities $\alpha=5/2$, 3, and 4 in  $D=2+1$ dimensions. We also comment on the demonstration of the Fierz-Pauli constraints for the general  case of arbitrary spin (helicity).  
	\end{abstract}
	\newpage
	
	\tableofcontents

         \newpage
	
	\section{Introduction}
	
	Understanding the fundamental physical laws is central to theoretical physics. At the core of this understanding lies the exploration of symmetries and symmetry groups, which are essential for characterizing the interactions and describing elementary particles. In this regard, the Poincaré Group plays a crucial role by unifying Lorentz transformations with spacetime translations.
	
	The seminal works by Wigner and Bargmann \cite{wigner, bargmann e wigner} have significantly shaped our understanding of the relationship between symmetries and the properties of elementary particles, such as mass and spin. These elementary particles belong to the unitary irreducible representations of the Poincaré Group labeled by the eigenvalues of the Casimir operators of the algebra, forming the mathematical basis for the description of elementary particles \cite{weinberg}. 
	
	Additionally, for a massive particle, the so-called Fierz-Pauli constraints reduce the degrees of freedom of a tensor (or spinor-tensor) representation to match the physical degrees of freedom. These constraints were postulated long ago in \cite{fp}.
	
	In the present article  we work both in $D=3+1$ and $D=2+1$ spacetime dimensions and focus on deriving explicit representations for the Poincaré group generators acting on tensors and spinor-tensors for arbitrary integer and half-integer spins (and helicities), respectively. Explicit expressions in the literature can only be found for specific lower spins  as $s\le 2$ in $D=3+1$, see \cite{giacosa}, and lower helicities in $D=2+1$  as $\alpha=3/2, 2$ and $3$, see \cite{elias e diego nge spin-3/2, denis e elias mestra spin-2, denis e elias nge spin-3}. 
 
       We also demonstrate the Fierz-Pauli constraints for massive particles directly from the eigenvalue equations of the Casimir invariants in a model-independent way for the higher spin (HS) cases $\mathfrak{s}=5/2,3,4$ in $D=3+1$ and higher helicity cases $\mathfrak{\alpha}=5/2,3,4$ in $D=2+1$. The Fierz-Pauli constraints are usually obtained in a dynamical way through the manipulation of the equations of motion of a given model. Here, differently, we demonstrate that they can be obtained from the algebra, generalizing \cite{giacosa}, where only the simpler lower spin cases $\mathfrak{s}=1/2,1, 3/2,2$ have been investigated. In $D=2+1$ our demonstrations for $\mathfrak{\alpha}=5/2 $ and $\mathfrak{\alpha}=4$ are new.

	Massive HS particles of arbitrary integer and half-integer spins are present in the spectrum of the superstring theory. Nowadays, motivations for HS theories are much broader, though, see, for instance, \cite{book_hs,snowmass_22}. We can speculate that string theory itself might be a spontaneously broken phase of a higher spin gravity (HiSGRA) theory that generalizes  Einstein's  spin-2 gravity, see \cite{rakibur string} and references therein for reviewing these motivations. The powerful higher spin local symmetries of HiSGRA would contribute to a possible finiteness of the corresponding quantum field theory, see the seminal earlier works \cite{fradkin,vasiliev}. Furthermore, there are also generalizations of the AdS/CFT duality to HiSGRA \cite{ss} and HS  applications to $3D$ bosonization \cite{mz,li}.  
	
	Even beyond bosonization, the case of $D=2+1$ dimensions is especially interesting since, differently from $D=3+1$, helicity eigenstates can be described by local field theories in $D=2+1$, which are called self-dual models or parity singlets. Interestingly, the helicity $+\alpha$ and helicity $-\alpha$ parity singlets can be ``soldered'' into a parity doublet containing both helicities $\pm \alpha$. These soldered models are the usual massive spin-$\mathfrak{s}$ models in any dimension $D\ge 3$ for $\mathfrak{s}=1, 3/2, 2$. Therefore, the parity singlets in $D=2+1$ work like the basic building blocks of massive particles in the real world in $D=3+1$. 

	This article is structured as follows: Section~\ref{sec:iso3} examines the Poincaré Group in $D=3+1$ spacetime, $ISO(3,1)$. We obtain general explicit expressions for the group generators within representations of arbitrary-rank tensors and spinor-tensors. Next, Section~\ref{sec:iso2} extends these results to the Poincaré Group in $D=2+1$ spacetime, $ISO(2,1)$, where we then identify explicit representations of the Casimir operators of the corresponding algebraic structure $\mathfrak{iso}(2,1)$. Notably, these results can be compared with the findings of Jackiw and Nair, as presented in \cite{jackiw e nair}, see also the precursor work on the Poincaré group in $D=2+1$ by Binegar \cite{binegar}. Then, in Section~\ref{sec:fp}, we obtain the Fierz-Pauli constraints for massive spin(helicity)-$\mathfrak{s}(\alpha)$ ($\mathfrak{s}(\alpha)=5/2,3,4$) particles in $D=3+1$  $(D=2+1)$. For the convenience of our readers,  in the appendices, we introduce explicit expressions of symmetric spin generators in $D=2+1$ dimensions. 
\\
 	
	{\bf{Conventions:}} In both $D=3+1$ and $D=2+1$, the Minkowski metric $\eta_{\mu\nu}$ is mostly plus, i.e., $\eta = \text{diag}(-,+,+,+)$ and $\eta = \text{diag}(-,+,+)$, respectively. Arbitrary tensors and spinor-tensors, with rank-$s$, depending on the spacetime coordinates $x^{\mu}$, are represented by $T^{\mu_{1}...\mu_{s}}(x)$ and $\psi^{\mu_{1}...\mu_{s}}(x)$, respectively. We use the word ``spin" and the symbol $\mathfrak{s}$ for  $D=3+1$ dimensions and ``helicity" and the symbol $\alpha$ for   $D=2+1$. Throughout the text, groups are referred to using capital letters while gothic letters are employed to represent algebras.	Symmetrizations, denoted by ``(...)," and anti-symmetrizations, represented by ``[...]," are normalized. 
	
	The following properties of the Dirac matrices are useful. In $D=2+1$, they are $2\times 2$ real matrices expressed in terms of Pauli matrices, as follows: $\gamma^{0} = i\sigma_2$, $\gamma^1 = \sigma_2$, and $\gamma^2 = \sigma_3$. These matrices satisfy the following fundamental relations:
	\bea \left\lbrack\gamma_\alpha,\gamma_\beta\right\rbrack&=&2\epsilon_{\alpha\beta\rho}\gamma^\rho,\\
	\left\lbrace\gamma_{\mu},\gamma_{\nu}\right\rbrace&=&2\eta_{\mu\nu},\\ 
	\gamma_{\mu}\gamma_{\nu}&=&\eta_{\mu\nu}+\epsilon_{\mu\nu\alpha}\gamma^{\alpha}. \eea
	
	\section{Poincaré Group $ISO(3,1)$}\label{sec:iso3}
	
	\hspace{0.5cm}It is well known that relativistic classical and quantum field theories must be invariant under the Poincaré transformations, which act on a four-vector $x^\mu$ spacetime coordinate as: 
	\be x^\mu\longrightarrow x'^\mu=\Lambda^\mu_{\s\nu}x^\nu+a^\mu\quad, \label{pt} \ee
	i.e., such transformation is a composition of the Lorentz transformation parameterized by $\Lambda^\mu_{\s\nu}$, $SO(3,1)$, plus a spacetime translation parameterized by $a^\mu$, which together form the inhomogeneous Lorentz group $ISO(3,1)$. Here, $a^\mu$ is a real four-vector while $\Lambda^\mu_{\s\nu}$ is a rank-two tensor that fulfills the relation:
	
	\be \eta_{\mu\nu}\Lambda^\mu_{\s\alpha}\Lambda^\nu_{\s\beta}=\eta_{\alpha\beta}\quad . \label{lt condition} \ee
	
	Since the Poincaré group is a Lie group,\footnote{A generic element of a Lie group, say $g(\theta)$, where $\theta$ is a continuous infinitesimal parameter, can always be represented by $D_R(g(\theta))=e^{i\theta_aT^{a}_R}$, for a given representation $D_R$, where $a=1,..., n$, while $T_R$ are the generators of the group corresponding to the representation $D_R$. This property is independent of the representation. For a pedagogical review, see for instance \cite{m.magiore}.} we can express Poincaré group transformations as
	\be \Lambda^\mu_{\s\nu}=\delta^\mu_\nu-\omega^\mu_{\s\nu},\quad a^\mu=\epsilon^\mu\quad, \label{infinitesimal lt} \ee
	where $\epsilon^\mu$ and $\omega^\mu_{\s\nu}$ are continuous infinitesimal parameters. In order to satisfy the condition (\ref{lt condition}), the tensor $\omega^\mu_{\s\nu}$ must be antisymmetric, i.e., $\omega^\mu_{\s\nu}=-\omega_\mu^{\s\nu}$. Therefore, in $D=3+1$ dimensions, we have six independent components corresponding to three rotation angles plus three Lorentz boosts. Using (\ref{infinitesimal lt}), the transformation (\ref{pt}) will be given by:
	\be  x^\mu\longrightarrow x'^\mu=x^\mu-\omega^\mu_{\s\nu}x^\nu+\epsilon^\mu\quad. \label{infinitesimal pt} \ee
	
    The generators of the corresponding Lie algebra satisfy the commutation relations: 
	\bea&[I^{\alpha\beta},I^{\mu\nu}]=&i(\eta^{\alpha\mu}I^{\beta\nu}+\eta^{\beta\nu}I^{\alpha\mu}-\eta^{\alpha\nu}I^{\beta\mu}-\eta^{\beta\mu}I^{\alpha\nu}),\\
	&[P^\mu,I^{\alpha\beta}]=&i(\eta^{\mu\beta}P^\alpha-\eta^{\mu\alpha}P^\beta),\\
	&[P^\alpha,P^\beta]=&0,\eea
	where $I^{\alpha\beta}=-I^{\beta\alpha}$ denotes the six generators of the Lorentz group associated with the six independent components of the antisymmetric parameter $\omega_{\mu\nu}$, and $P^\alpha$ denotes the four generators of spacetime translations associated with the four parameters $\epsilon_{\mu}$.

	\subsection{Arbitrary rank-$s$ tensors in $D=3+1$}
	\hspace{0.5cm}In this section, we obtain explicit expressions for the generators of the Poincaré group acting on arbitrary tensors of rank-$s$, for integer s.  We consider translations and Lorentz transformations separately in order to obtain the generators, $P^\alpha$ and $I^{\alpha\beta}$, respectively. We start with the translation case. 
	
	\subsubsection{Spacetime translations}\label{sec:trans_3}
	
	Considering an infinitesimal spacetime translation:
	\be x'^\mu=x^\mu+\epsilon^\mu, \label{translation}\ee 
	
	\no Since $\p x^{\prime\mu}/\p x^{\nu} = \delta^{\mu}_{\nu} $ we have:
	\be T'^{\mu_{1}...\mu_{s}}(x')=T^{\mu_{1}...\mu_{s}}(x)\quad. \label{Ts} \ee
	On the one hand, expanding the left hand side of (\ref{Ts}),  we obtain:
	\be T'^{\mu_{1}...\mu_{s}}(x)=T^{\mu_{1}...\mu_{s}}(x)-\epsilon^\alpha\p_\alpha T^{\mu_{1}...\mu_{s}}(x) +\mathcal{O}(\epsilon^2) \quad,\label{t'exp}\ee
	On the other hand, as a group element acting on the original tensor, we have:
	\be T'^{\mu_{1}...\mu_{s}}(x)=\left\lbrack exp\left(-i\epsilon_\alpha P^\alpha\right)\right\rbrack ^{\mu_{1}...\mu_{s}}_{\s\nu_{1}...\nu_{s}}T^{\nu_{1}...\nu_{s}}(x)\quad. \ee
	By expanding the exponential on the right-hand side,  we obtain:
	\be T'^{\mu_{1}...\mu_{s}}(x)=T^{\mu_{1}...\mu_{s}}(x)-i\epsilon_\alpha(P^\alpha)^{\mu_{1}...\mu_{s}}_{\s \nu_{1}...\nu_{s}}T^{\nu_{1}...\nu_{s}}(x)+\mathcal{O}(\epsilon^2)\quad.\label{t'group}\ee 
	Comparing (\ref{t'exp}) with (\ref{t'group}), we finally get an explicit expression for the generator of translations:
	\be (P^\alpha)^{\mu_{1}...\mu_{s}}_{\s\nu_{1}...\nu_{s}}=-i\left(\prod\limits_{i=1}^s\delta^{\mu_i}_{\nu_i}\right)\p^\alpha\quad,\label{P4} \ee
	which is valid on arbitrary rank-$s$ tensors in $D=3+1$ dimensions.\\

	\subsubsection{Lorentz transformations}

	In order to obtain the explicit expressions for the generators of the Lorentz transformations $I^{\alpha\beta}$, we use the same approach of the previous case, but now, notice that under the Lorentz transformation $\p x^{\prime\mu}/\p x^{\nu}  = \Lambda^{\mu}_{\nu} $, so we have:
	\be  T'^{\mu_{1}...\mu_{s}}(x')=\left(\prod\limits_{i=1}^s\Lambda^{\mu_i}_{\s\nu_i}\right)T^{\nu_{1}...\nu_{s}}(x)\quad.\label{tl field} \ee  
	In the infinitesimal case, considering that $\Lambda^{\mu}_{\s\nu}= \delta^{\mu}_{\s\nu}- \omega^{\mu}_{\s\nu} $  up until the first order in $\omega$, we have explicitly:
	\bea \prod\limits_{i=1}^s\Lambda^{\mu_i}_{\s\nu_i}&=&(\delta^{\mu_1}_{\nu_1}\delta^{\mu_2}_{\nu_2}...\delta^{\mu_s}_{\nu_s}-\omega^{\mu_1}_{\nu_1}\delta^{\mu_2}_{\nu_2}\delta^{\mu_3}_{\nu_3}...\delta^{\mu_s}_{\nu_s}-\omega^{\mu_2}_{\nu_2}\delta^{\mu_1}_{\nu_1}\delta^{\mu_3}_{\nu_3}...\delta^{\mu_s}_{\nu_s}
	-\omega^{\mu_s}_{\nu_s}\delta^{\mu_1}_{\nu_1}...\delta^{\mu_{s-1}}_{\nu_{s-1}}\delta^{\mu_s}_{\nu_s})\nonumber\\
	&=&\prod\limits_{i=1}^s\delta^{\mu_i}_{\nu_i}-\sum\limits_{k=1}^s\prod\limits_{i=1\atop i\neq k}^s\delta^{\mu_i}_{\nu_i}\omega^{\mu_k}_{\s\nu_k}\quad.\label{tl exp}\eea 
	Inserting (\ref{tl exp}) into (\ref{tl field}) and expanding the left hand side of (\ref{tl field}) up to the first order in $\omega$, we have:
	\be T'^{\mu_{1}...\mu_{s}}(x)-\omega^\gamma_{\s\alpha}x^\alpha\p_\gamma T^{\mu_{1}...\mu_{s}}(x)=\left[\prod\limits_{i=1}^s\delta^{\mu_i}_{\nu_i}-\sum\limits_{k=1}^s\prod\limits_{i=1\atop i\neq k}^s\delta^{\mu_i}_{\nu_i}\omega^{\mu_k}_{\s\nu_k}\right]T^{\nu_{1}...\nu_{s}}(x)\quad,\ee
	making explicit use of the antisymmetry of $\omega_{\alpha\beta}\equiv \eta_{\alpha\gamma}\omega^{\gamma}_{\s\beta}$ we have:
	\be T'^{\mu_{1}...\mu_{s}}(x)=T^{\mu_{1}...\mu_{s}}(x)+\frac{\omega_{\alpha\beta}}{2}\left[(x^\beta\p^\alpha-x^\alpha\p^\beta)\prod\limits_{i=1}^s\delta^{\mu_i}_{\nu_i}-\sum\limits_{k=1}^s\prod\limits_{i=1\atop i\neq k}^s\delta^{\mu_i}_{\nu_i}(\eta^{\alpha\mu_k}\delta^\beta_{\nu_k}-\eta^{\beta\mu_k}\delta^\alpha_{\nu_k})\right]T^{\nu_{1}...\nu_{s}}(x)\label{t'l exp}\ee
	As an element of the Lorentz group acting on a rank-$s$ tensor, we have:
	\bea T'^{\mu_{1}...\mu_{s}}(x)&=&exp\Big(-\frac{i}{2}\omega_{\alpha\beta}I^{\alpha\beta}\Big)^{\mu_{1}...\mu_{s}}_{\s\nu_{1}...\nu_{s}}T^{\nu_{1}...\nu_{s}}(x)\quad, \nn\\
	&=&T^{\mu_{1}...\mu_{s}}(x)-\frac{i}{2}\omega_{\alpha\beta}(I^{\alpha\beta})^{\mu_{1}...\mu_{s}}_{\s\nu_{1}...\nu_{s}}T^{\nu_{1}...\nu_{s}}(x) + {\cal O}(\omega^2)\quad,\label{tl group} \eea
where $I^{\alpha\beta}$ is the generator of Lorentz transformations which we are looking for. Comparing (\ref{tl group}) with (\ref{t'l exp}), we get an expression for the generators of Lorentz transformation:
	\be (I^{\alpha\beta})^{\mu_{1}...\mu_{s}}_{\s\nu_{1}...\nu_{s}}=i\left(\prod\limits_{i=1}^s\delta^{\mu_i}_{\nu_i}\right)(x^\beta\p^\alpha-x^\alpha\p^\beta)-i\sum\limits_{k=1}^s\prod\limits_{i=1\atop i\neq k}^s\delta^{\mu_i}_{\nu_i}(\eta^{\alpha\mu_k}\delta^\beta_{\nu_k}-\eta^{\beta\mu_k}\delta^\alpha_{\nu_k})\label{Lorentz G}\quad.\ee
	This generator can be decomposed into an orbital angular momentum part  and a spin part $I^{\alpha\beta} = L^{\alpha\beta} + S^{\alpha\beta} $, explicitly,
	\bea (L^{\alpha\beta})^{\mu_{1}...\mu_{s}}_{\s\nu_{1}...\nu_{s}}&=&i\prod\limits_{i=1}^s\delta^{\mu_i}_{\nu_i}(x^\beta\p^\alpha-x^\alpha\p^\beta)\quad,\label{LT}\\
	(S^{\alpha\beta})^{\mu_{1}...\mu_{s}}_{\s\nu_{1}...\nu_{s}}&=&i\sum\limits_{k=1}^s\prod\limits_{i=1\atop i\neq k}^s\delta^{\mu_i}_{\nu_i}(\eta^{\beta\mu_k}\delta^\alpha_{\nu_k}-\eta^{\alpha\mu_k}\delta^\beta_{\nu_k})\quad.\label{ST}\eea
	Now, we have all the generators of the Poincaré group on rank-$s$ tensors, with $s$ integer in $D=3+1$ dimensions. Next, we generalize such results to the fermionic case represented by arbitrary rank-$s$ spinor-tensors.
	
	\subsection{Arbitrary rank-$s$ spinor-tensors in $D=3+1$}
	\hspace{0.5cm}In this section we obtain the generators of the Poincaré group on rank-$s$ spinor-tensors. We denote spinor-tensors as $\psi^{\mu_1\mu_2...\mu_s}(x)$. Below, we derive the expressions of the generators proceeding as before. 
	
	\subsubsection{Spacetime translations and Lorentz transformations}
	
	\hspace{0.5cm} We start by requiring that the translated spinor-tensor at the translated spacetime coordinate is equal to the original spinor-tensor at the original spacetime coordinate:
	\be \psi'^{\mu_{1}...\mu_{s}}(x')=\psi^{\mu_{1}...\mu_{s}}(x)\quad,\label{field ts} \ee
	If we follow the same steps of Section~\ref{sec:trans_3},  we end up with the following translation generator on rank-s spinor tensors:
	\be (P^\alpha)^{\mu_{1}...\mu_{s}}_{\s\nu_{1}...\nu_{s}}=-i\prod\limits_{i=1}^s\delta^{\mu_i}_{\nu_i}\p^\alpha\mathbb{I}\quad.\label{P4S} \ee
	\no where $\mathbb{I}$ is a ($4\times4$) identity matrix in the Dirac-spinor space. Except by the matrix $\mathbb{I}$, we have exactly the same result obtained in the pure tensor case (\ref{P4}).
	
	Now, regarding the Lorentz transformations, there is a difference with respect to the bosonic case. Namely, the Lorentz transformation of a rank-s spinor tensor is given by
	\be \psi'^{\mu_{1}...\mu_{s}}(x')=\left(\prod\limits_{i=1}^s\Lambda^{\mu_{1}...\mu_{s}}_{\s\nu_{1}...\nu_{s}}\right)\,\,S\,\,\psi^{\nu_{1}...\nu_{s}}(x)\quad, \label{lts}\ee
	where we take into account i) the tensor representation of the Lorentz group through the factors $\Lambda$ acting directly on the Minkowski indices, and ii) the spinorial representation given by $S$, which is explicitly given by the group element:
    \be S\equiv exp\left(\frac{i}{4}\omega_{\alpha\beta}\sigma^{\alpha\beta}\right)\quad,\label{st} \ee
	with the generators $\sigma_{\alpha\beta}$ given by
 \be  \sigma_{\alpha\beta}=-\frac{i}{2}\left[\gamma_\alpha,\gamma_\beta\right]\quad \ee
	Notice that, in (\ref{st}), the $\omega_{\alpha\beta}$ parameters are precisely the same infinitesimal antisymmetric parameters of the Lorentz transformations. However, the generators $\sigma_{\alpha\beta}$ are the corresponding Lorentz generators of the spinorial counterpart ($\sigma_{\alpha\beta}$ is explicitly antisymmetric). The symbol $\left [\,\,, \,\,\right]$ stands for commutator while the $\gamma's$ are the usual Dirac matrices. 
	
	On the one hand, expanding the left hand side of (\ref{lts}) and (\ref{st}), using (\ref{tl exp}), we have:
\be \psi'^{\mu_{1}...\mu_{s}}(x)-\omega^\gamma_{\s\alpha}x^\alpha\p_\gamma \psi'^{\mu_{1}...\mu_{s}}(x)=\left[\prod\limits_{i=1}^s\delta^{\mu_i}_{\nu_i}-\sum\limits_{k=1}^s\prod\limits_{i=1\atop i\neq k}^s\delta^{\mu_i}_{\nu_i}\omega^{\mu_k}_{\s\nu_k} + \cdots\right]\left[\mathbb{I}+\frac{i}{4}\omega_{\alpha\beta}\sigma^{\alpha\beta}+ \cdots\right]\psi^{\nu_{1}...\nu_{s}}(x)\, ..\
\ee
	\no Recalling the antisymmetry of $\omega_{\alpha\beta}$, we have:
	\bea \psi'^{\mu_{1}...\mu_{s}}(x)&=&\psi^{\mu_{1}...\mu_{s}}(x)+\frac{\omega_{\alpha\beta}}{2}\left[( x^\beta\p^\alpha-x^\alpha\p^\beta)\prod\limits_{i=1}^s\delta^{\mu_i}_{\nu_i}\mathbb{I} \right.
	\nn\\
	 &-& \left. \sum\limits_{k=1}^s\prod\limits_{i=1\atop i\neq k}^s\delta^{\mu_i}_{\nu_i}(\eta^{\alpha\mu_k}\delta^\beta_{\nu_k}-\eta^{\beta\mu_k}\delta^\alpha_{\nu_k})\mathbb{I}    
	+\frac{i}{2}\sigma^{\alpha\beta}\prod\limits_{i=1}^s\delta^{\mu_i}_{\nu_i}\right]\psi^{\nu_{1}...\nu_{s}}(x).\label{s'l exp}\eea
 On the other hand,
	\bea  \psi'^{\mu_{1}...\mu_{s}}(x)&=&exp\Big(-\frac{i}{2}\omega_{\alpha\beta}I^{\alpha\beta}\Big)^{\mu_{1}...\mu_{s}}_{\s\nu_{1}...\nu_{s}}\psi^{\nu_{1}...\nu_{s}}(x)\nn\\ &=&\psi^{\mu_{1}...\mu_{s}}(x)-\frac{i}{2}\omega_{\alpha\beta}(I^{\alpha\beta})^{\mu_{1}...\mu_{s}}_{\s\nu_{1}...\nu_{s}}\psi^{\nu_{1}...\nu_{s}}(x)\quad. \label{tlsg} \eea
	Comparing (\ref{tlsg}) with (\ref{s'l exp}), the generator for Lorentz transformations can be written as:
	\be (I^{\alpha\beta})^{\mu_{1}...\mu_{s}}_{\s\nu_{1}...\nu_{s}}=i\prod\limits_{i=1}^s\delta^{\mu_i}_{\nu_i}(x^\beta\p^\alpha-x^\alpha\p^\beta)\mathbb{I}-i\sum\limits_{k=1}^s\prod\limits_{i=1\atop i\neq k}^s\delta^{\mu_i}_{\nu_i}(\eta^{\alpha\mu_k}\delta^\beta_{\nu_k}-\eta^{\beta\mu_k}\delta^\alpha_{\nu_k})\mathbb{I}-\frac{1}{2}\sigma^{\alpha\beta}\prod\limits_{i=1}^s\delta^{\mu_i}_{\nu_i}\quad. \label{Lorentz G S}\ee
	Again, this generator can be decomposed into an orbital angular momentum part and a spin part $I^{\alpha\beta} = L^{\alpha\beta} +  S^{\alpha\beta}$ respectively given by: 
	\bea (L^{\alpha\beta})^{\mu_{1}...\mu_{s}}_{\s\nu_{1}...\nu_{s}}&=&i\prod\limits_{i=1}^s\delta^{\mu_i}_{\nu_i}(x^\beta\p^\alpha-x^\alpha\p^\beta)\mathbb{I},\label{LS}\\
	(S^{\alpha\beta})^{\mu_{1}...\mu_{s}}_{\s\nu_{1}...\nu_{s}}&=&-i\sum\limits_{k=1}^s\prod\limits_{i=1\atop i\neq k}^s\delta^{\mu_i}_{\nu_i}(\eta^{\alpha\mu_k}\delta^\beta_{\nu_k}-\eta^{\beta\mu_k}\delta^\alpha_{\nu_k})\mathbb{I}-\frac{1}{2}\sigma^{\alpha\beta}\prod\limits_{i=1}^s\delta^{\mu_i}_{\nu_i}\label{SS},\eea
	where in the spin part $S^{\alpha\beta}$ of the total angular momentum we have an explicit separation of the integer -- the first term of the right-hand side of (\ref{SS})-- as well as the semi-integer contributions -- the second term of the right-hand side of (\ref{SS}). 
 
 We have, therefore, obtained representations of all Poincaré generators for arbitrary rank-$s$ tensors and spinor-tensors. Next, we study the Casimir operators for this group.

	\subsection{Casimir operators of ${\mathfrak{iso}}(3,1)$}

The Casimir operators of a group are the operators that commute with all group generators. Consequently, these operators are invariant by the group symmetry transformations and are used to classify the irreducible representations associated with that group. According to the seminal work by Wigner in 1939 \cite{wigner}, mass and spin are precisely the two properties that characterize systems with invariance under the Poincaré group. Then, the irreducible representation of such a group is the basis of the description of elementary particles.

	The Lie algebra ${\mathfrak{iso}}(3,1)$ of the Poincaré group is composed of the generators $P^{\mu}$ and $I^{\alpha\beta}$.  In $D=3+1$, it is well known that the Poincaré algebra has two Casimirs,\footnote{From now on, we call a Casimir operator simply Casimir.}
	\bea C_{1}\equiv P^{2} &=& P_{\alpha}P^{\alpha},\label{c1}\\
	C_{2}\equiv W^{2} &=& W_{\alpha}W^{\alpha}= -\frac{1}{2}I_{\alpha\beta}I^{\alpha\beta}P^{2}+I_{\gamma\beta}I^{\alpha\beta}P_{\alpha}P^{\gamma} \label{c2}, \eea
	where $W^{\alpha}$ is  the Pauli-Lubanski pseudovector, given explicitly by 
	\be W^{\alpha}=\epsilon^{\alpha\beta\gamma\delta}P_{\beta}I_{\gamma\delta}/2\label{W},\ee
	\no and $I_{\gamma\delta}$ is given by (\ref{Lorentz G}) (for tensors) or (\ref{Lorentz G S}) (for spinor-tensors).
	The corresponding eigenvalue equations,
	\bea  P^{2}\phi &=& -m^2\phi\quad , \label{ee1} \\ W^{2}\phi &=& -m^{2}\mathfrak{s}(\mathfrak{s}+1)\phi \quad  \label{ee2}, \eea
	define the mass ($m$) and the spin ($\mathfrak{s}$) of an elementary massive particle represented (suppressing indices) by a field $\phi$.
	
	For the arbitrary rank-$s$ tensor representation, we explicitly write the Casimir $C_1$ with the representation given by (\ref{P4}). Once the eigenvalue equation (\ref{ee1}) must be satisfied, we obtain the Klein-Gordon equation in the given representation:
	\be (\square - m^2)T^{\mu_1...\mu_s}=0.\ee
	Similarly, we write the eigenvalue equation of the Casimir $C_2$ using  (\ref{P4}) and (\ref{Lorentz G}) to get
	\bea
	& &2\, s \,\square T^{\mu_{1}...\mu_{s}} + 2 \sum_{\overset{k,r=1}{k < r}}^{s} \produtoria_{\overset{i=1}{i\neq k,r}}^{s} \delta^{\mu_i}_{\nu_i}\delta^{\mu_k}_{\nu_r}\delta^{\mu_r}_{\nu_k}\square T^{\nu_1...\nu_s} - 2\sum_{k=1}^{s} \produtoria_{\overset{i=1}{i\neq k}}^{s} \delta^{\mu_i}_{\nu_i}\del_{\nu_k}\del^{\mu_k}T^{\nu_1...\nu_s} + \nn \\
	&+& 2 \sum_{\overset{k,r=1}{k < r}}^{s} \produtoria_{\overset{i=1}{i\neq k,r}}^{s} \delta^{\mu_i}_{\nu_i}\left[\eta^{\mu_k\mu_r}\del_{\nu_k}\del_{\nu_r}-\eta_{\nu_k\nu_r}(\square\eta^{\mu_k\mu_r}-\del^{\mu_k}\del^{\mu_r})-\delta^{\mu_r}_{\nu_k}\del^{\mu_k}\del_{\nu_r}-\delta^{\mu_k}_{\nu_r}\del^{\mu_r}\del_{\nu_k} \right]T^{\nu_1...\nu_s}\nn\\
	&=&m^2 \mathfrak{s}(\mathfrak{s}+1)T^{\mu_1...\mu_s}.
	\eea
	Analogously, on arbitrary rank-$s$ spinor-tensors, the eigenvalue equations are respectively:
	
	\be (\square - m^2)\psi^{\mu_1...\mu_s}=0,\ee
	\bea
	&&\Big(2s+\frac{3}{4}\Big) \square \psi^{\mu_{1}...\mu_{s}} + 2 \sum_{\overset{k,r=1}{k < r}}^{s} \prod_{\overset{i=1}{i\neq k,r}}^{s}\delta^{\mu_i}_{\nu_i}\delta^{\mu_k}_{\nu_r}\delta^{\mu_r}_{\nu_k}\square \psi^{\nu_1...\nu_s} - 2\sum_{k=1}^{s} \prod_{\overset{i=1}{i\neq k}}^{s}\delta^{\mu_i}_{\nu_i}\p_{\nu_k}\p^{\mu_k}\psi^{\nu_1...\nu_s} + \nonumber \\
	&+& 2 \sum_{\overset{k,r=1}{k < r}}^{s} \prod_{\overset{i=1}{i\neq k,r}}^{s} \delta^{\mu_i}_{\nu_i}\left[\eta^{\mu_k\mu_r}\p_{\nu_k}\p_{\nu_r}-\eta_{\nu_k\nu_r}(\square\eta^{\mu_k\mu_r}-\p^{\mu_k}\p^{\mu_r})-\delta^{\mu_r}_{\nu_k}\p^{\mu_k}\p_{\nu_r}-\delta^{\mu_k}_{\nu_r}\p^{\mu_r}\p_{\nu_k} \right]\psi^{\nu_1...\nu_s}\nonumber\\
	&-&\frac{1}{2}\sum_{k=1}^{s}\prod_{\overset{i=1}{i\neq k}}^{s}\delta^{\mu_i}_{\nu_i}([\cancel{\p},\gamma_{\nu_k}]\p^{\mu_k}-[\cancel{\p},\gamma^{\mu_k}]\p_{\nu_k}-[\gamma^{\mu_k},\gamma_{\nu_k}]\Box)\psi^{\nu_1...\nu_s}\nonumber\\
	&=&m^2 \mathfrak{s}(\mathfrak{s}+1)\psi^{\mu_{1}...\mu_{s}},\label{c2s2+1}
	\eea
	where $\cancel{\p}$  denotes $\gamma_\mu\p^\mu$.

 	Now that we have all the generators of the Poincaré group and the eigenvalue equations of the Casimir operators for rank-$s$ tensors and spinor-tensors in $D=3+1$, we proceed with a simple prescription that allows us to adapt our results to the lower dimension case of the $D=2+1$ spacetime.
	
	\section{Poincaré group $ISO(2,1)$}\label{sec:iso2}
	
	We start by taking advantage of that, in three spacetime dimensions, the 3 independent components of an antisymmetric tensor of rank two $I^{\mu\nu}$ can be written in terms of a three-component vector $I_{\alpha}$ as
	\be I_\alpha=\frac{1}{2}\epsilon_{\alpha\mu\nu}I^{\mu\nu}\quad.\label{transcription}  \ee
	Then, the corresponding commutators of the Poincaré algebra $\mathfrak{iso}(2,1)$ are  
	\bea &[I_\mu,I_\nu]=&i\epsilon_{\mu\nu\rho}I^\rho\quad, \\
	&[I_\mu,P_\nu]=&i\epsilon_{\mu\nu\rho}P^\rho\quad, \\
	&[P_\mu,P_\nu]=&0\quad. \eea
	
	\subsection{Arbitrary rank-$s$ tensors and spinor-tensors in $D=2+1$}
	
	We easily verify that the generators for spacetime translation for tensors (\ref{P4}) and spinor-tensors (\ref{P4S}) in $D=3+1$ preserve their expressions in $D=2+1$. 
	
	The orbital angular momentum and the spin part are therefore
	\be (L_\rho)^{\mu_{1}...\mu_{s}}_{\s\nu_{1}...\nu_{s}}=\frac{1}{2}\epsilon_{\rho\alpha\beta}(L^{\alpha\beta})^{\mu_{1}...\mu_{s}}_{\s\nu_{1}...\nu_{s}}\quad, \ee
	\be (S_\rho)^{\mu_{1}...\mu_{s}}_{\s\nu_{1}...\nu_{s}}=\frac{1}{2}\epsilon_{\rho\alpha\beta}(S^{\alpha\beta})^{\mu_{1}...\mu_{s}}_{\s\nu_{1}...\nu_{s}}\quad, \ee
	so that the generators (\ref{LT}) and (\ref{ST}) are generalized respectively to:
	\bea (L_\rho)^{\mu_{1}...\mu_{s}}_{\s\nu_{1}...\nu_{s}}&=&-i\epsilon_{\rho\alpha\beta}x^\alpha\p^\beta\prod\limits_{i=1}^s\delta^{\mu_i}_{\nu_i}\label{LB3}\quad,\\
	(S_\rho)^{\mu_{1}...\mu_{s}}_{\s\nu_{1}...\nu_{s}}&=&i\sum\limits_{k=1}^s\prod\limits_{i=1\atop i\neq k}^s\epsilon^{\mu_k}_{\s\rho\nu_k}\delta^{\mu_i}_{\nu_i}\quad.\label{SB3}
	\eea
These are the representations of the generator of Lorentz transformations for rank-$s$ tensors in $D=2+1$, describing bosonic particles of mass $m$ and helicity $\alpha$.\footnote{Throughout this work we use the word helicity to refer to the ``spin'' of the particle in $D=2+1$. Notice that, in $D=3+1$, massive spin-${\mathfrak{s}}$ particles propagate $2{\mathfrak{s}}+1$ degrees of freedom, while, in $D=2+1$, they propagate helicity $+\alpha$ {\bf and} $-\alpha$ in the case of parity preserving models (doublets of parity) and $+\alpha$ {\bf or} $-\alpha$ in the case of parity breaking models (singlets of parity described by self-dual descriptions).}
	
	Proceeding in the very same way for the  fermionic case, with generators (\ref{LS}) and (\ref{SS}), we obtain:
	\bea (L_\rho)^{\mu_{1}...\mu_{s}}_{\s\nu_{1}...\nu_{s}}&=&-i\epsilon_{\rho\alpha\beta}x^\alpha\p^\beta\prod\limits_{i=1}^s\delta^{\mu_i}_{\nu_i}\mathbb{I}\label{LF3}\quad,\\
	(S_\rho)^{\mu_{1}...\mu_{s}}_{\s\nu_{1}...\nu_{s}}&=&i\sum\limits_{k=1}^s\prod\limits_{i=1\atop i\neq k}^s\epsilon^{\mu_k}_{\s\rho\nu_k}\delta^{\mu_i}_{\nu_i}\mathbb{I}+\frac{i}{2}\prod\limits_{i=1}^s\delta^{\mu_i}_{\nu_i}\gamma_\rho\quad,\label{SF3}
	\eea
	which are the representations of the corresponding generators of the Lorentz transformations on spinor-tensors in $D = 2 + 1$. Next, we explore a little more the Casimirs of the algebra ${\mathfrak{iso}}(2,1)$ and their eigenvalue equations.

	\subsection{Casimirs of ${\mathfrak{iso}}(2,1)$} 
	As mentioned in the introduction, the classification of the unitary irreducible representations of the Poincaré group in three dimensions was for the very first time provided by Binegar in $1981$ \cite{binegar}. An interesting application of this study was given by Jackiw and Nair in 1991 \cite{jackiw e nair}, in which they studied one-particle states as unitary representations of the Poincaré group in three dimensions, aiming the description of {\it anyons}. In section II of \cite {jackiw e nair}, the authors provide the Poincaré algebra in $D=2+1$ dimensions and its Casimirs:
 
 \begin{itemize}
     \item $C_1$: $P^2$, satisfying the eigenvalue equation $(P^2+m^2) \phi=0$ for a given one particle state $\phi$, where $m$ denotes the mass of a given one particle state represented by $\phi$.

     \item $C_2$: $P\cdot S$, the $D=2+1$ version of the Pauli-Lubanski operator,  satisfying $(P \cdot S - \alpha \, m)\phi=0$, where $\alpha$ specifies the helicity of the particle.
 \end{itemize}  
 
 In \cite{jackiw e nair}, the authors explored spin-1/2 and spin-1 representations for $ISO(2,1)$. In the present section, we generalize such results for arbitrary rank-$s$ tensors and spinor-tensors in the same way we have studied for $ISO(3,1)$. The eigenvalue equations are:
	
	\bea C_1 \phi &\equiv& P^\mu P_\mu\phi=-m^2\phi\label{c1_2+1}\quad,\\
	C_2 \phi &\equiv&  P^\mu S_\mu\phi=\alpha \, m \phi\label{c2_2+1}\quad, \eea

	\no where $S_{\mu}$ is given by (\ref{SB3}) or  (\ref{SF3}). For arbitrary rank-$s$ tensors, the explicit expressions for the eigenvalue equations (\ref{c1_2+1}) and (\ref{c2_2+1}) are respectively given by:
	
	\be (\Box-m^2)T^{\mu_{1}...\mu_{s}}=0\quad , \ee
	\be  \left(\sum\limits_{k=1}^s\prod\limits_{i=1\atop i\neq k}^sE^{\mu_k}_{\s\nu_k}\delta^{\mu_i}_{\nu_i}\right)T^{\s\nu_{1}...\nu_{s}}=\alpha\, m \,T^{\mu_{1}...\mu_{s}}\quad. \label{PLB}\ee 
	where we have defined the operator $E^\mu_{\s\nu}\equiv\epsilon^{\mu\rho}_{\s\s\nu}\p_\rho$. 
	The equation (\ref{PLB}) is the Pauli-Lubanski equation for an arbitrary rank-$s$ field in the tensor representation.
	
	Analogously,  in the spinor-tensor representation, using  (\ref{P4S}) and (\ref{SF3}), we have:
	\be \left(\Box -m^2\right)\psi^{\mu_{1}...\mu_{s}}=0\label{kgs}\quad,
	\ee
	\be
	\left(\sum\limits_{k=1}^s\prod\limits_{i=1\atop i\neq k}^sE^{\mu_k}_{\s\nu_k}\delta^{\mu_i}_{\nu_i}\mathbb{I}+\frac{1}{2}\prod\limits_{i=1}^s\delta^{\mu_i}_{\nu_i}\gamma_\rho\p^\rho\right)\psi^{\nu_{1}...\nu_{s}}-\alpha \, m \psi^{\mu_{1}...\mu_{s}}=0\quad.\label{PLF} 
	\ee
	In the next section, we demonstrate that it is possible to obtain the Fierz-Pauli constraints from the Casimir eigenvalue equations.
	
		\section{Fierz-Pauli constraints from the algebra}\label{sec:fp}
	
	As we increase the spin of the massive particle, in order to propagate the correct number of physical degrees of freedom in a given representation, higher-rank tensors or spinor-tensors are required. However, higher-rank tensors introduce spurious field components that must be eliminated. This is possible by introducing constraint equations. For massive particles of integer spin-$\mathfrak{s}$, represented by arbitrary rank-$s$ bosonic fields, those constraints are given by the Fierz-Pauli constraints \cite{fp}:
	\bea T^{\mu_1...\mu_s}&=&T^{(\mu_1...\mu_s)},\\
	\eta_{\mu_1\mu_2}T^{\mu_1\mu_2...\mu_s}&=&0,\\
	\p_{\mu_1}T^{\mu_1\mu_2...\mu_s}&=&0.\eea
    For semi-integer spin, represented by arbitrary rank-s fermionic fields, they are:
	\bea  \psi^{\mu_1...\mu_s}&=&\psi^{(\mu_1...\mu_s)},\label{simf}\\
	\eta_{\mu_1\mu_2}\psi^{\mu_1\mu_2...\mu_s}&=&0,\\
	\p_{\mu_1}\psi^{\mu_1\mu_2...\mu_s}&=&0,\\
	\gamma_{\mu_1}\psi^{\mu_1\mu_2...\mu_s}&=&0.\label{gtf}\eea
	
	The derivation of Fierz-Pauli constraints from a given massive higher-spin model is typically not a straightforward task, as it requires working through the equations of motion in order to demonstrate that the constraints hold and all lower-rank auxiliary fields vanish on-shell. However, in \cite{giacosa}, the authors have achieved a model-independent derivation of the Fierz-Pauli constraints for spins 1/2, 1, 3/2, and 2 in $D=3+1$. In the present work, we extend these results to include bosonic cases of spin-3 and spin-4 particles, along with the fermionic example of spin-5/2 in $D=3+1$. We also explore the generalization  of our findings to $D=2+1$ for helicities $5/2,3,4$  and pave the way for a demonstration for arbitrary integer spin (helicity). 

\subsection{Spin-5/2}

 In this section, we obtain the Fierz-Pauli constraints for a rank-2 spinor-tensor representation of a spin-5/2 massive particle in \dtres and \ddoiS, from the eigenvalue equations provided by the Casimirs $C_1$ and $C_2$.
	
	\subsubsection{Spin-5/2 in \dtres:}
 
For a rank-2 spinor-tensor, the first Casimir operator $C_1$ has the Klein-Gordon equation as its eigenvalue equation:
	\be \Box\psi^{\mu\nu}=m^2\psi^{\mu\nu}\label{c1-5/2}. \ee
	Setting $s=2$ and $\mathfrak{s}=5/2$ in (\ref{c2s2+1}) and making use of the Klein-Gordon equation (\ref{c1-5/2}), we find after some manipulation, the  following expression for the Pauli-Lubaski equation:
	\bea
	&&m^2(-2\psi^{\mu\nu}-\eta^{\mu\nu}\psi+\psi^{\nu\mu})+\p^\mu\p^\nu\psi+\eta^{\mu\nu}\p_\alpha\p_\beta\psi^{\alpha\beta}+\nn\\
	&-&\p^\mu\p_\alpha\psi^{\alpha\nu}-\p^{\nu}\p_\beta\psi^{\mu\beta}-\p^\mu\p_\beta\psi^\nu\beta-\p^\nu\p_\alpha\psi^{\alpha\mu}+\nn\\
	&-&\frac{1}{4}([\cancel{\p},\gamma_\alpha]\p^{\mu}-[\cancel{\p},\gamma^{\mu}]\p_\alpha-m^2[\gamma^\mu,\gamma_\alpha])\psi^{\alpha\nu}+\nn\\
	&-&\frac{1}{4}([\cancel{\p},\gamma_\beta]\p^{\nu}-[\cancel{\p},\gamma^{\nu}]\p_\beta-m^2[\gamma^\nu,\gamma_\beta])\psi^{\mu\beta}=0,\label{PL-5/2d3}
	\eea
	where $\psi$ denote $\eta_{\mu\nu}\psi^{\mu\nu}$.

The following contractions will result in the Fierz-Pauli constraints:
	\bea \p^2(\ref{PL-5/2d3})&\Rightarrow &\p_\mu\p_\nu\psi^{\mu\nu}=0,\\
	\gamma\p(\ref{PL-5/2d3})&\Rightarrow &\gamma_\nu\p_\mu\psi^{\mu\nu}=\gamma_\mu\p_\nu\psi^{\mu\nu}=0,\\
	\p(\ref{PL-5/2d3})&\Rightarrow &\p_\mu\psi^{\mu\nu}=\p_\nu\psi^{\mu\nu}=0,\label{5/2d}\\
	\eta(\ref{PL-5/2d3})&\Rightarrow &\psi=0,\label{5/2t}\\
	\gamma\gamma(\ref{PL-5/2d3})_a&\Rightarrow &\gamma_\mu\gamma_\nu\psi^{\mu\nu}_a=0,\label{as1}\\
	\gamma(\ref{PL-5/2d3})_a&\Rightarrow &\gamma_\mu\psi^{\mu\nu}_a=\gamma_\nu\psi_a^{\mu\nu}=0\Rightarrow\psi_{a\mu\nu}=0,\label{as2}\\
	\gamma\gamma(\ref{PL-5/2d3})&\Rightarrow &\gamma_\mu\gamma_\nu\psi^{\mu\nu}=0,\\
	\gamma(\ref{PL-5/2d3})&\Rightarrow &\gamma_\mu\psi^{\mu\nu}=\gamma_\nu\psi^{\mu\nu}=0.\label{5/2gt}\eea
	
	In the equations (\ref{as1}) and (\ref{as2}) the subscript $``a"$ means that we are taking only the antisymmetric part of $\psi^{\mu\nu}$, i.e.,  $\psi^{\mu\nu}_a=(\psi^{\mu\nu}-\psi^{\nu\mu})/2$.
	
	With the equations   (\ref{5/2d}), (\ref{5/2t}), (\ref{as2}) and (\ref{5/2gt}) in hands, we complete the demonstration of the Fierz-Pauli constraints (\ref{simf}) - (\ref{gtf}) for a rank-2 spinor-tensor in \dtres, obtained from the Casimir equations. 
	
	\subsubsection{Helicity-5/2 in \ddoiS}
	For a rank-2 spinor-tensor with helicity $\alpha=5/2$, the expression (\ref{kgs}) yields the Klein-Gordon equation
	while (\ref{PLF}) provides the Pauli-Lubanski equation in \ddoiS:
	\be E^\mu_{\s\rho}\psi^{\rho\nu}+E^{\nu}_{\s\rho}\psi^{\mu\rho}+\frac{1}{2}\cancel{\p}\psi^{\mu\nu}=\frac{5}{2}m\psi^{\mu\nu}.\label{PL-5/2}\ee
	First, applying $\cancel{\p}$ to equation (\ref{PL-5/2}) and using (\ref{PL-5/2}) itself in the terms with $\cancel{\p}\psi^{\alpha\beta}$, we have:
	\be 9m^2\psi^{\mu\nu}+2\p^\mu\p_\lambda\psi^{\lambda\nu}+2\p^\nu\p_\lambda\psi^{\mu\lambda}-E^\mu_{\s\rho}E^\nu_{\s\lambda}\psi^{\rho\lambda}=5m\cancel{\p}\psi^{\mu\nu\label{PL-5/2.2}}.\ee
	Then, as we have done in $D=3+1$, we suggest some steps in order to obtain the Fierz-Pauli constraints, but now it is always necessary to make use of the Klein-Gordon equation and equations (\ref{PL-5/2}) and (\ref{PL-5/2.2}). Combining these equations, we have:
	\bea 
	\p^2(\ref{PL-5/2});\p^2(\ref{PL-5/2.2})&\Rightarrow &\p_\mu\p_\nu\psi^{\mu\nu}0,\\
	\eta(\ref{PL-5/2});\eta(\ref{PL-5/2.2})&\Rightarrow &\eta_{\mu\nu}\psi^{\mu\nu}=\label{5/2-2}0,\\
	\p\gamma(\ref{PL-5/2});\p\gamma(\ref{PL-5/2.2})&\Rightarrow & \p_\mu\gamma_\nu\psi^{\mu\nu}=\p_\nu\gamma_\mu\psi^{\mu\nu}=0,\\
	\gamma^2(\ref{PL-5/2});\gamma^2(\ref{PL-5/2.2})&\Rightarrow & \gamma_\mu\gamma_\nu\psi^{\mu\nu}=\gamma_\nu\gamma_\mu\psi^{\mu\nu}=0,\\
	\p(\ref{PL-5/2});\p(\ref{PL-5/2.2})&\Rightarrow &\p_\mu\psi^{\mu\nu}=\p_\nu\psi^{\mu\nu}=0,\label{5/2-1}\\
	\gamma(\ref{PL-5/2});\gamma(\ref{PL-5/2.2})&\Rightarrow &\gamma_\mu\psi^{\mu\nu}=\gamma_\nu\psi^{\mu\nu}=0.\label{5/2-3}
	\eea
	Notice that after obtaining the transversality condition and using the property $E_{\mu\nu}E_{\alpha\beta}=\Box(\theta_{\mu\beta}\theta_{\nu\alpha}-\theta_{\mu\alpha}\theta_{\nu\beta})$ the equation (\ref{PL-5/2.2}) reduces to a Dirac equation:
	\be(\cancel{\p}-m)\psi^{\mu\nu}=0.\ee
	In order to demonstrate that $\psi^{\mu\nu}$ is symmetric, we apply $\epsilon_{\mu\nu\alpha}$ to (\ref{PL-5/2}), which gives us:
	\be \epsilon_{\mu\nu\alpha}\psi^{\mu\nu}=0\Rightarrow \psi^{\mu\nu}=\psi^{\nu\mu}.\label{5/2 ant}\ee
With this result and the conditions (\ref{5/2-2}), (\ref{5/2-1}), and (\ref{5/2-3}) the Fierz-Pauli constraints are all verified from the Casimir equations.

	\subsection{Spin-3}

  Next, we obtain the Fierz-Pauli constraints for a rank-3 tensor representation of a spin-3 massive particle in \dtres and \ddoiS, from the Casimir equations.

  \subsubsection{Spin-3 in $D=3+1$}
	
The Casimirs $C_1$ and $C_2$ are given respectively by (\ref{c1})  and (\ref{c2}),
	\be C_{1}\equiv P^{2} = P_{\alpha}P^{\alpha} \Rightarrow (\Box - m^{2})T^{\mu\nu\gamma} = 0,
	\label{casimir p2 spin3}\ee
	\be C_{2}\equiv W^{2} = W_{\alpha}W^{\alpha}\Rightarrow W^{2}T^{\mu\nu\gamma}=m^{2}\mathfrak{s}(\mathfrak{s}+1) T^{\mu\nu\gamma} = 12 m^2 T^{\mu\nu\gamma}.
	\label{casimir w2 spin3}\ee
	
	Using (\ref{W}) explicitly in the expression (\ref{casimir w2 spin3}) and  (\ref{casimir p2 spin3}), we find after some manipulation:
	\begin{align}
		&6m^{2}T^{\mu\nu\gamma}+2m^{2}(\eta^{\nu\gamma}T^{\mu\sigma}{}_{\sigma}+\eta^{\mu\gamma}T^{\sigma\nu}{}_{\sigma}+\eta^{\mu\nu}T^{\sigma}{}_{\sigma}{}^{\gamma})-2m^2(T^{\mu\gamma\nu}+T^{\gamma\nu\mu}+T^{\nu\mu\gamma})+\nn \\
		&+2(\del_{\alpha}\del^{\gamma}T^{\mu\nu\alpha}+\del_{\alpha}\del^{\nu}T^{\mu\alpha\gamma}+\del_{\alpha}\del^{\mu}T^{\alpha\nu\gamma})-2(\eta^{\nu\gamma}\del_{\alpha}\del_{\sigma}T^{\mu\alpha\sigma}+\eta^{\mu\gamma}\del_{\alpha}\del_{\sigma}T^{\alpha\nu\sigma}+\eta^{\mu\nu}\del_{\alpha}\del_{\sigma}T^{\alpha\sigma\gamma})+ \nn \\
		&-2\del^{\nu}\del^{\gamma}T^{\mu\sigma}{}_{\sigma}-2\del^{\mu}\del^{\gamma}T^{\sigma\nu}{}_{\sigma}-2\del^{\mu}\del^{\nu}T^{\sigma}{}_{\sigma}{}^{\gamma}+2\del_{\alpha}\del^{\gamma}T^{\mu\alpha\nu}+2\del_{\alpha}\del^{\gamma}T^{\alpha\nu\mu}+\nn \\
		&+2\del_{\alpha}\del^{\nu}T^{\mu\gamma\alpha}+2\del_{\alpha}\del^{\nu}T^{\alpha\mu\gamma}+2\del_{\alpha}\del^{\mu}T^{\nu\alpha\gamma}+2\del_{\alpha}\del^{\mu}T^{\gamma\nu\alpha} = 0. 
		\label{PauliLubanski spin3}
	\end{align}
	
	Applying $\eta_{\nu\gamma}$, $\eta_{\mu\nu}$, and $\eta_{\mu\gamma}$, respectively, on (\ref{PauliLubanski spin3}), and taking one derivative of the resulting equations, we show that 
	\be 
	T^{\mu\sigma}{}_{\sigma}=0 \quad\quad;\quad\quad T^{\sigma}{}_{\sigma}{}^{\gamma}=0 \quad\quad;\quad\quad T^{\sigma\nu}{}_{\sigma}=0.
	\label{traceless spin3}
	\ee
	Inserting the results (\ref{traceless spin3}) in (\ref{PauliLubanski spin3}) and applying $\partial_{\mu}\partial_{\nu}\partial_{\gamma}$ in the obtained expression we demonstrate that $\partial_{\mu}\partial_{\nu}\partial_{\gamma}T^{\mu\nu\gamma}=0$. Next, we apply $\partial_{\mu}\partial_{\gamma}$, $\partial_{\mu}\partial_{\nu}$, and $\partial_{\nu}\partial_{\gamma}$, respectively, on (\ref{PauliLubanski spin3}) to show that $\partial_{\mu}\partial_{\gamma}T^{\mu\nu\gamma}=0$, $\partial_{\mu}\partial_{\nu}T^{\mu\nu\gamma}=0$, and $\partial_{\nu}\partial_{\gamma}T^{\mu\nu\gamma}=0$. Finally, applying $\partial_{\mu}$, $\partial_{\nu}$, and $\partial_{\gamma}$, respectively, we get the transversality conditions:
	\be 
	\del_{\mu}T^{\mu\nu\gamma}=0 \quad\quad;\quad\quad \del_{\nu}T^{\mu\nu\gamma}=0 \quad\quad;\quad\quad \del_{\gamma}T^{\mu\nu\gamma}=0.
	\label{transversalidade spin3}
	\ee
	Then, the expression (\ref{PauliLubanski spin3}) can be rewritten as: 
	\be 
	3\, T^{\mu\nu\gamma}=T^{\mu\gamma\nu}+T^{\nu\mu\gamma}+T^{\gamma\nu\mu}.
	\label{identity spin3}
	\ee 
	\no By  taking differences of   (\ref{identity spin3}) and relabelling some indices we obtain the following system
	\bea 
	4\, t_1^{\mu\nu\gamma} - t_2^{\gamma\nu\mu}-  t_3^{\mu\gamma\nu} &=& 0, \nn \\  -\, t_1^{\mu\nu\gamma} +4\, t_2^{\gamma\nu\mu}+  t_3^{\mu\gamma\nu} &=& 0, \label{s3}\\ -\, t_1^{\mu\nu\gamma} +\, t_2^{\gamma\nu\mu}+  4\, t_3^{\mu\gamma\nu} &=& 0, \nn \eea 
	\no  where $t_1^{\mu\nu\gamma} \equiv T^{\mu\nu\gamma} - T^{\nu\mu\gamma}$, $t_2^{\gamma\nu\mu} \equiv T^{\gamma\nu\mu} - T^{\gamma\mu\nu}$ and  $t_3^{\mu\gamma\nu} \equiv T^{\nu\gamma\mu} - T^{\mu\gamma\nu}$. Since the numerical matrix corresponding to the linear system (\ref{s3}) has a non-vanishing determinant, the only solution of (\ref{s3}) is the trivial one: $t_j^{\mu\nu\rho} =0$, $ \, j=1,2,3$. Consequently, we end up with a totally symmetric tensor
	\be T^{\mu\nu\gamma}=T^{(\mu\nu\gamma)}. \label{sym3}\ee

	In this way, we deduce the Fierz-Pauli constraints for rank-3 tensors, i.e., the tensor is fully symmetric (\ref{sym3}), traceless (\ref{traceless spin3}) and transverse (\ref{transversalidade spin3}). These constraints reduce the degrees of freedom of the arbitrary tensor $T^{\mu\nu\gamma}$ from 64 to 7, in agreement with the $2\mathfrak{s}+1$ rule for the number of degrees of freedom of a massive spin-$\mathfrak{s}$  particle in $D=3+1$ at $\mathfrak{s}=3$.
	
	\subsubsection{Helicity-3 in $D=2+1$}
	For a rank-3 tensor with helicity $\alpha=3$, the Pauli-Lubanski equation (\ref{PLB}) is given by:
	\bea		 
	E_{\gamma}{}^{\rho}T^{\mu\nu\gamma}+E_{\beta}{}^{\nu}T^{\mu\beta\rho}+E_{\alpha}{}^{\mu}T^{\alpha\nu\rho}=3mT^{\mu\nu\rho}.
	\label{AD3}
	\eea
	Using the Klein-Gordon equation in (\ref{AD3}), we can demonstrate the Fierz-Pauli constraints after some manipulations. We summarize the steps as follows:
	\bea
	\p^3(\ref{AD3})&\Rightarrow &\p_\mu\p_\nu\p_\rho T^{\mu\nu\rho}=0\\
	\p^2(\ref{AD3})&\Rightarrow &\p_\mu\p_\nu T^{\mu\nu\rho}=\p_\mu\p_\rho T^{\mu\nu\rho}=\p_\rho\p_\nu T^{\mu\nu\rho}=0\\
	\p(\ref{AD3})&\Rightarrow &\p_\mu T^{\mu\nu\rho}=\p_\nu T^{\mu\nu\rho}=\p_\rho T^{\mu\nu\rho}=0\\
	\eta(\ref{AD3})&\Rightarrow & T^{\mu\s\rho}_{\s\mu}=T^{\mu\nu}_{\s\s\mu}=T^{\mu\nu}_{\s\s\nu}=0
	\eea
	where, repeated use of (\ref{AD3}) has been done. Contracting $E^\gamma_{\s\mu},E^\gamma_{\s\nu}$ and $E^\gamma_{\s\rho}$,  one by one, with the (\ref{AD3}) and reinserting the three resulting equations in (\ref{AD3}), we obtain: 
	\be 3\, T^{\mu\nu\gamma}=T^{\mu\gamma\nu}+T^{\nu\mu\gamma}+T^{\gamma\nu\mu}\ee
	that is the same equation (\ref{identity spin3}), so performing the same steps of the previous section we conclude that the tensor is totally symmetric. This completes the demonstration of the Fierz-Pauli constraints for rank-3 tensors in $D=2+1$.

 	\subsection{Spin-4}

  Here we obtain the Fierz-Pauli constraints for a rank-4 tensor representation of a spin-4 massive particle in \dtres and \ddoiS, from the Casimir equations.
	
	\subsubsection{Spin-4 in $D=3+1$}
	As usual, the first Casimir operator $C_1$ yields the Klein-Gordon equation:
	\be 
	\square T^{\mu\nu\alpha\beta} = m^2 \;T^{\mu\nu\alpha\beta}.
	\label{KG4}
	\ee
	
	\no In turn, for $s=4$ and $\mathfrak{s}=4$, we rewrite the Pauli-Lubanski eigenvalue equation in the form:	
	\be 
	m^2 A^{\mu\nu\alpha\beta} + B^{\mu\nu\alpha\beta} = 0, 
	\label{eqspin4}
	\ee
	\no where 
	\be 
	A^{\mu\nu\alpha\beta}= 6 \; T^{\mu\nu\alpha\beta} - (T^{\nu\mu\alpha\beta}+T^{\alpha\nu\mu\beta}+T^{\beta\nu\alpha\mu}+T^{\mu\alpha\nu\beta}+T^{\mu\beta\alpha\nu}+T^{\mu\nu\beta\alpha}),
	\ee
	\no while
	\begin{align}
		B^{\mu\nu\alpha\beta}&= \del^{\mu}\del_{\gamma}T^{\gamma\nu\alpha\beta}+\del^{\mu}\del_{\gamma}T^{\nu\gamma\alpha\beta}+\del^{\mu}\del_{\gamma}T^{\alpha\nu\gamma\beta}+\del^{\mu}\del_{\gamma}T^{\beta\nu\alpha\gamma}+\del^{\nu}\del_{\gamma}T^{\gamma\mu\alpha\beta}+\del^{\nu}\del_{\gamma}T^{\mu\gamma\alpha\beta}\nn \\
		&+\del^{\nu}\del_{\gamma}T^{\mu\alpha\gamma\beta}
		+\del^{\nu}\del_{\gamma}T^{\mu\beta\alpha\gamma}+\del^{\alpha}\del_{\gamma}T^{\gamma\nu\mu\beta}+\del^{\alpha}\del_{\gamma}T^{\mu\gamma\nu\beta}+\del^{\alpha}\del_{\gamma}T^{\mu\nu\gamma\beta}+\del^{\alpha}\del_{\gamma}T^{\mu\nu\beta\gamma}\nn \\
		&+\del^{\beta}\del_{\gamma}T^{\gamma\nu\alpha\mu}+\del^{\beta}\del_{\gamma}T^{\mu\gamma\alpha\nu}+\del^{\beta}\del_{\gamma}T^{\mu\nu\gamma\alpha}+
		\del^{\beta}\del_{\gamma}T^{\mu\nu\alpha\gamma}\nn\\
		&-\del_\gamma\del_\sigma(\eta^{\mu\nu}T^{\gamma\sigma\alpha\beta}+\eta^{\mu\alpha}T^{\gamma\nu\sigma\beta}+\eta^{\mu\beta}T^{\gamma\nu\alpha\sigma}+\eta^{\nu\alpha}T^{\mu\gamma\sigma\beta}+\eta^{\nu\beta}T^{\mu\gamma\alpha\sigma}
		+\eta^{\alpha\beta}T^{\mu\nu\gamma\sigma})\nn\\
		&-m^2(\theta^{\mu\nu}T^{\gamma}{}_{\gamma}{}^{\alpha\beta}+\theta^{\mu\alpha}T^{\gamma\nu}{}_{\gamma}{}^{\beta}+\theta^{\mu\beta}T^{\gamma\nu\alpha}{}_{\gamma}+\theta^{\nu\alpha}T^{\mu\gamma}{}_{\gamma}{}^{\beta}+\theta^{\nu\beta}T^{\mu\gamma\alpha}{}_{\gamma}+\theta^{\alpha\beta}T^{\mu\nu\gamma}{}_{\gamma}). 
		\label{Bmnab}
	\end{align}
	
	In order to verify the Fierz-Pauli constraints, we perform a series of contractions and operations starting from (\ref{eqspin4}), always relying on (\ref{KG4}). Due to the large number of contraction possibilities offered by the four indices of the tensor, the required procedure for the demonstration is laborious and typically involves the use of systems of equations that are obtained from different manipulations of (\ref{eqspin4}). We summarize the steps as follows:
	\bea
	\del^{4}\,\, (\ref{eqspin4}) &\Rightarrow& \del_{\mu}\del_{\nu}\del_\alpha\del_\beta T^{\mu\nu\alpha\beta} = 0\\
	\del^{3}\,\,(\ref{eqspin4}) &\Rightarrow& \del_{\mu}\del_{\nu}\del_\alpha T^{\mu\nu\alpha\beta} = \del_{\mu}\del_{\nu}\del_\beta T^{\mu\nu\alpha\beta} = \del_{\mu}\del_\alpha\del_\beta T^{\mu\nu\alpha\beta} =\del_{\nu}\del_\alpha\del_\beta T^{\mu\nu\alpha\beta}= 0\\
	\del^2\,\eta\,\, (\ref{eqspin4}) &\Rightarrow& \del_{\mu}\del_{\nu} T^{\mu\nu\gamma}{}_{\gamma}  = \del_{\mu}\del_{\nu} T^{\mu\gamma\nu}{}_{\gamma}= 0 = \cdots \\
	\del^2\,\, (\ref{eqspin4}) &\Rightarrow& \del_{\mu}\del_\nu T^{\mu\nu\alpha\beta} = \del_{\mu}\del_\nu T^{\mu\alpha\nu\beta} = 0 = \cdots  \\
	\del\, \eta\,\, (\ref{eqspin4}) &\Rightarrow& \del_{\mu}T^{\mu\beta\gamma}{}_{\gamma}=\del_{\mu}T^{\mu\gamma\beta}{}_{\gamma} = \del_{\mu}T^{\beta\mu\gamma}{}_{\gamma} =0= \cdots  \\
	\eta\,\eta\,\,(\ref{eqspin4}) &\Rightarrow& T^{\gamma}{}_{\gamma}{}^\alpha{}_\alpha= T^{\gamma\alpha}{}_{\gamma\alpha}= T^{\gamma\alpha}{}_{\alpha\gamma}=0  \\
	\eta\,\,(\ref{eqspin4}) &\Rightarrow& T^{\gamma}{}_{\gamma}{}^{\alpha\beta}=T^{\alpha\beta\gamma}{}_{\gamma}=0=\cdots\label{t} \label{tr}  \\
	\del\,\, (\ref{eqspin4}) &\Rightarrow& \del_{\mu}T^{\mu\nu\alpha\beta}=\del_{\mu}T^{\nu\mu\alpha\beta}=0=\cdots, \label{dt} 	\eea
	where, in each step, we feedback equation (\ref{eqspin4}) and after that it becomes:
	\be 
	6 \; T^{\mu\nu\alpha\beta} = T^{\nu\mu\alpha\beta} + T^{\alpha\nu\mu\beta}+T^{\beta\nu\alpha\mu}+T^{\mu\alpha\nu\beta}+T^{\mu\beta\alpha\nu}+T^{\mu\nu\beta\alpha}.
	\label{simetrico4}
	\ee
	\no Similarly to the spin-3 case, it is convenient to define six partially antisymmetric tensors:
	\bea t_1^{\mu\nu\alpha\beta} = T^{\mu\nu\alpha\beta}-T^{\nu\mu\alpha\beta} \quad  &;&  \quad  t_4^{\alpha\mu\nu\beta} = T^{\alpha\mu\nu\beta}-T^{\alpha\nu\mu\beta}  \nn\\
	t_2^{\nu\alpha\mu\beta} = T^{\nu\alpha\mu\beta}-T^{\mu\alpha\nu\beta} \quad  &;&  \quad  t_5^{\beta\mu\alpha\nu} = T^{\beta\mu\alpha\nu}-T^{\beta\nu\alpha\mu}\quad \label{t4j}\\ t_3^{\nu\beta\alpha\mu} = T^{\nu\beta\alpha\mu} -T^{\mu\beta\alpha\nu} \quad  &;&  \quad  t_6^{\beta\alpha\mu\nu} = T^{\beta\alpha\mu\nu}-T^{\beta\alpha\nu\mu}  \nn \eea
	By taking differences of  (\ref{simetrico4}) we derive, for instance,
	\be
	7 t_1^{\mu\nu\alpha\beta}+t_1^{\mu\nu\beta\alpha}+t_2^{\nu\alpha\mu\beta}+t_3^{\nu\beta\alpha\mu} +  t_4^{\alpha\mu\nu\beta}+ t_5^{\beta\mu\alpha\nu} = 0.
	\label{antissimetrico4}
	\ee
	\no Notice that, in principle, $t_1^{\mu\nu\alpha\beta}\ne t_1^{\mu\nu\beta\alpha}$. Therefore, the exchange $\alpha \leftrightarrow \beta $ in (\ref{t4j}) leads to a total of 12 partially antisymmetric rank-4 tensors. Taking differences of  (\ref{simetrico4}) and relabelling indices we can derive 12 equations for those 12 tensors. As in the spin-3 case, the determinant of the numerical $12 \times 12$  matrix of the corresponding linear system is different than zero, and the trivial solution $t_j^{\gamma\delta\rho\lambda}=0$ for $j=1,2, \cdots, 12$ is the only possible one. This immediately implies that the tensor is symmetric under the exchange of any pair of indices, 
	
	\be T^{\gamma\delta\rho\lambda}=T^{(\gamma\delta\rho\lambda)} \quad . \label{sym4} \ee
	
	\no Thus, together with (\ref{tr}) and (\ref{dt}), we have demonstrated the Fierz-Pauli constraints for rank-4 tensors in $D=3+1$, obtained from the Casimir eigenvalue equations.

	\subsubsection{Helicity-4 in $D=2+1$}
	For an arbitrary rank-4 tensor, the equation (\ref{PLB}) is given by:
	\be
	E_{\lambda}{}^{\mu}T^{\lambda\nu\alpha\beta}+E_{\lambda}{}^{\nu}T^{\mu\lambda\alpha\beta}+E_{\lambda}{}^{\alpha}T^{\mu\nu\lambda\beta}+E_{\lambda}{}^{\beta}T^{\mu\nu\alpha\lambda} = 4\;m \;T^{\mu\nu\alpha\beta}.
	\label{helicidade4}
	\ee
	
	Similarly to what was done in \dtres, we will describe the main steps in the search for the Fierz-Pauli constraints, namely: 
	\bea
        \p^3\,\, (\ref{helicidade4}) &\Rightarrow&  \p_\mu\p_\nu\p_\alpha\p_\beta T^{\mu\nu\alpha\beta} = 0 \\
	\del^3\,\, (\ref{helicidade4}) &\Rightarrow& \del_\mu\del_\nu\del_\alpha T^{\mu\nu\alpha\beta} = 0\\
	\del^2\, \eta\,\, (\ref{helicidade4}) &\Rightarrow& \del_\mu\del_\nu T^{\mu\nu\gamma}{}_{\gamma} = 0\\
	E\,\del^2\,\, (\ref{helicidade4}) &\Rightarrow& \del_\mu\del_\nu T^{\mu\nu\alpha\beta} = 0\\
	\del\, \eta\,\, (\ref{helicidade4}) &\Rightarrow& \del_\mu T^{\mu\nu\gamma}{}_{\gamma} = 0\\
	E\,\eta\,\eta\,\, (\ref{helicidade4}) &\Rightarrow& T^{\mu}{}_{\mu}{}^{\gamma}{}_{\gamma} = 0\\
	E\,\eta\,\,(\ref{helicidade4}) &\Rightarrow& T^{\mu\nu\gamma}{}_{\gamma} = 0\\ 
	E\,\del\,\, (\ref{helicidade4})&\Rightarrow& \del_\mu T^{\mu\nu\alpha\beta} = 0
	\eea
	It is worth mentioning that we always use the Klein-Gordon equation $\square T^{\mu\nu\alpha\beta}=m^2 T^{\mu\nu\alpha\beta}$ whenever it is possible as well as the equation (\ref{helicidade4}) itself is reused. We have the following steps
	Contracting $E_{\mu}{}^{\rho}$, $E_{\nu}{}^{\rho}$, $E_{\alpha}{}^{\rho}$, and $E_{\beta}{}^{\rho}$ with (\ref{helicidade4}) and reinserting the obtained equations into (\ref{helicidade4}), we have:
	\be 
	6 \; T^{\mu\nu\alpha\beta} = T^{\nu\mu\alpha\beta} + T^{\alpha\nu\mu\beta}+T^{\beta\nu\alpha\mu}+T^{\mu\alpha\nu\beta}+T^{\mu\beta\alpha\nu}+T^{\mu\nu\beta\alpha}
	\ee
	which is the same equation (\ref{simetrico4}) obtained in \dtreS. This last result demonstrates, therefore, that the tensor is symmetric with respect to any pair of indices. We have then demonstrated that the Fierz-Pauli constraints for a rank-4 tensor also follow from the eigenvalue equations of the Casimir operators of the Poincaré group in \ddoiS.

	\subsection{Comments on the case of arbitrary rank-s tensors}\label{sec:higher-rank}

	We have observed that in the $D=3+1$, for example in the case of spin-4, only the part $A^{\mu\nu\alpha\beta}$ survived in (\ref{eqspin4}) after deducing the transversality and tracelless conditions. From this, we conclude that the tensor is completely symmetric. In general, for arbitrary rank-s, we would obtain:
	\be m^2 A^{\mu_1...\mu_s} + B^{\mu_1...\mu_s} =0. \label{caso geral} \ee
	
	The procedure to be followed when we are demonstrating the Fierz-Pauli constraints is then to show that $B^{\mu_1...\mu_s} = 0$. To do this, we must apply $s$ derivatives to (\ref{caso geral}) in order to demonstrate that $\partial_{\mu_1}\dots\partial_{\mu_s}T^{\mu_1 \cdots \mu_s} = 0$. Then, we apply $(s-1)$ derivatives or $(s-2)$ derivatives plus the metric in order to eliminate terms of the form $\partial_{\mu_1}\dots\partial_{\mu_{s-1}}T^{\mu_1\cdots\mu_s}$ and $T^{\mu_1\cdots\gamma}{}_{\gamma}$.
	
	Notice that we first demonstrate that the scalars obtained from (\ref{caso geral}) are zero, and then the vectors and then rank-2 tensors, and so on. The idea is to gradually increase the rank of the objects. The difficulty arises because as the rank of the tensor increases, the number of possible contractions also increases, and the result is that the demonstration relies on increasingly larger systems of homogeneous equations.
	
	The ultimate goal is to demonstrate that the trace over an arbitrary pair of indices and a contracted derivative with some index of the tensor is zero. At this stage, we will be dealing with objects of rank $(s-2)$ and $(s-1)$ respectively. In this way, (\ref{caso geral}) reduces to $A^{\mu_1...\mu_s}=0$ which can be written as:
	
	\be
	\binom{s}{2} T^{\mu_1\mu_2\cdots\mu_s}= T^{\mu_2\mu_1\cdots\mu_s}+T^{\mu_3\mu_2\mu_1\cdots\mu_s}+\dots + T^{\mu_s\mu_2\cdots\mu_1}+\dots + T^{\mu_1\mu_2\cdots\mu_s\mu_{s-1}},
	\label{simetricoN}
	\ee
	
	\no where there are exactly $\binom{s}{2}$ terms on the right-hand side of the equation, corresponding to the simple interchange of two indices of the tensor $T^{\mu_1...\mu_s}$ on the left-hand side of the equation.
	
	To demonstrate that the tensor is completely symmetric, we must permute the indices of (\ref{simetricoN}), combining them to form antisymmetric tensors of the type $T^{[\mu_1\mu_2]\cdots\mu_s}$, as in equation (\ref{antissimetrico4}). Just as it was done for spin-4, we must construct $(s-2)!\binom{s}{2}=\frac{s!}{2}$ equations to account for the same number of variables (the number of ways to select two positions for the antisymmetric pair and then permute the remaining $(s-2)$ positions).	We then construct the matrix that represents this system\footnote{Strictly speaking, this is valid if the tensor indices are different. Otherwise, the variables will not be independent and the system will have its order reduced.} of homogeneous equations and calculate the determinant. It must be verified that the determinant is nonzero, so that the solution of the system is unique and trivial, meaning that each term $T^{[\mu\nu]\cdots\mu_s}$ is zero. This proves that the tensor is symmetric with respect to any pair of its indices. Therefore, it is concluded that the Pauli-Lubanski equation, together with the Klein-Gordon equation, both derived from first principles with the help of the eigenvalue equations of the Casimir operators of the Poincaré group, implies that the rank-$s$ tensor must satisfy the Fierz-Pauli constraints and thus has precisely $2\mathfrak{s}+1$ degrees of freedom.
	
	Some comments on the case of $D=2+1$ dimensions are also in order. we can notice that, in (\ref{AD3}) the operator $E$ acts on each of the indices of the tensor $T^{\mu\nu\rho}$. The same occurs in the equation for the rank-4 tensors. We conjecture that if we work with an arbitrary tensor of $s$ indices, we will have the equation:
	\be 
	E_{\alpha}{}^{\mu_1}T^{\alpha\mu_2\ldots\mu_s}+E_{\alpha}{}^{\mu_2}T^{\mu1\alpha\mu_3\ldots\mu_s}+\ldots E_{\alpha}{}^{\mu_s}T^{\mu_1\mu_2\ldots\mu_{(s-1)}\alpha}= \alpha\;m\;T^{\mu_1 \mu_2 \ldots \mu_s}.
	\label{ADk}
	\ee
	
	This is a necessary condition that tensors must satisfy in order to describe fields of helicity $+\alpha$ or $-\alpha$. However, it is not sufficient on its own as it does not provide an equation that is identical to the equation of motion of a massive self-dual model. It must be accompanied by the Klein-Gordon equation, which, in our work, is given by the first Casimir.

\section{Conclusions}
	
In this work, we have investigated the description of massive higher spin particles in $D=3+1$ and $D=2+1$ dimensions in a model-independent way. We derive the Fierz-Pauli constraints starting from algebraic properties of the Poincaré groups $ISO(3,1)$ and $ISO(2,1)$, respectively.

 The derivation of the Fierz-Pauli constraints from the eigenvalue equations of the Casimir operators of the Poincaré algebra seems to have been employed for the first time by \cite{giacosa}.  There, the authors have worked out specifically the spin cases $s=1/2, 1, 3/2$ and $s= 2 $ in $D=3+1$. They have conjectured that their results could be extended to higher spins. It is important to stress that in the higher spin cases ($s\ge 5/2$) the 
 corresponding massless models (Fronsdal and Fang-Fronsdal models \cite{Fronsdal:1978rb, Fang:1978wz})  are invariant only under constrained gauge transformations where the gauge parameters must be traceless (bosons) or $\gamma$-traceless (fermions). So one might expect eventually that the model-independent analysis carried out here for the massive cases with $s=5/2, 3$ and $s= 4 $ in $D=3+1$ would be affected by their massless counterparts and could present some surprises. We have shown that this is not the case and the  Fierz-Pauli constraints for the higher spin models are exactly the ones postulated in the literature.  

We have also extended the method of \cite{giacosa} to $D=2+1$ dimensions. More specifically,  in Sections~\ref{sec:iso3} and~\ref{sec:iso2}, we obtain explicit expressions for the translation generators, $P$, and the Lorentz rotation generators, $I=L+S$, acting on arbitrary rank tensors and spinor-tensors in both aforementioned dimensions. In each section, we also provide explicit expressions for the Casimirs of the algebras $\mathfrak{iso}(3,1)$, given by $C_1= P^2$ and $C_2=W^2$, and $\mathfrak{iso}(2,1)$, given by $C_1= P^2$ and $C_2=P\cdot S$.  With those expressions at hand and the respective eigenvalue equations defining mass and spin/helicity, in Section~\ref{sec:fp}, we demonstrate that the Fierz-Pauli constraints follow from the Casimir equations, independently of any specific model. Our results also justify the expressions for the Lorentz generators in $D=2+1$ used by \cite{gaitan}  and \cite{arias}, and complement previous works \cite{jackiw e nair,binegar}. Our specific results for helicities $5/2$ and $4$ are new.

From our expressions, we can deduce the Fierz-Pauli constraints that must be present in any tensor and spinor-tensor description of massive spinning particles. Regarding the general case of arbitrary spin, we have mentioned the technical difficulties involved and paved the way for a general demonstration, see subsection~\ref{sec:higher-rank}. Our expressions are also useful for higher rank descriptions. In particular, it is possible to confirm the constraints appearing in the model of  \cite{renato}, which describes massive spin-1 particles via a symmetric rank-2 tensor $W_{\mu\nu}=W_{\nu\mu}$. Such model accommodates self-interacting vertices with super-renormalizable couplings in $D=3+1$, like for example $V = g_o\,  (\eta^{\mu\nu}W_{\mu\nu})^3$ which can not be built up via usual vector or 
Kalb-Ramond anti-symmetric tensor descriptions. Moreover, we can also revisit the equations of motion of first-order self-dual models describing massive particles with helicity-1 \cite{townsend pilch e nieuwenhuizen}, 3/2 \cite{deser e kay}, 2 \cite{aragone spin-2}, 5/2 \cite{aragone spin-5/2}, 3, and 4 \cite{aragone spin-3 e spin-4} in $D=2+1$ dimensions in order to show that, after dynamically obtaining the Fierz-Pauli constraints, they turn into the Klein-Gordon plus Pauli-Lubanski equations in terms of the generators of the Poincaré algebra $\mathfrak{iso}(2,1)$. For convenience, in Appendix~\ref{app:A}, we provide the expressions for the spin generators for some of these cases. In appendix B we demonstrate that our original general expression for the spin operator for arbitrary helicity does satisfy the Lorentz algebra in $D=2+1$.

Additionally, our readers may also find interesting the works 
\cite{Lyakhovich:1998ij, Lyakhovich:2000zy} and the more recent ones  \cite{oh1,oh2} where explicit models
 for spinning particles at the level of first quantization (finite number of degrees of freedom) are obtained in different dimensions via the method of the coadjoint orbit of the Poincaré group.

\section*{Acknowledgements}

 HVAS is supported by FAPESP grant - 2022/14185-1.  DD is partially supported by CNPq  (grant 313559/2021-0). RRLdS is supported in part by the National Science Centre (Poland) under the research Grant No. 2020/38/E/ST2/00126.  DD and RRLdS acknowledge previous support by FAPESP grant 2016/09489-0. RRLdS thanks Astrid Eichhorn and her group at the University of Southern Denmark for their hospitality during the last stages of the work. The authors thank Prof. Julio M. Hoff da Silva for useful discussions.
	
\appendix 	
\section{Spin generators in $D=2+1$}\label{app:A}
	Here, for convenience of our readers we give explicit expressions for the spin generator for the specific cases of helicities  $\alpha=5/2, 3$ and $4$. With such expressions, we can better understand the equations of motion coming from the so-called first order in derivatives self-dual descriptions. The simpler cases of helicities $\alpha=1, 3/2$ and $2$ have been used, for example, in \cite{jackiw e nair},  \cite{denis e elias mestra spin-2} and \cite{elias e diego nge spin-3/2}, respectively.
	
\subsection{helicity - $\alpha=5/2$}

	The symmetric spin operator obtained from (\ref{SF3}), taking the rank $s=2$, is given by:
	\be(S_\rho)^{\mu\nu}_{\s\alpha\beta}=\frac{i}{2}(\delta^\mu_\alpha\epsilon^\nu_{\s\rho\beta}+\delta^\nu_\alpha\epsilon^\mu_{\s\rho\beta}+\delta^\mu_\beta\epsilon^\nu_{\s\rho\alpha}+\delta^\nu_\beta\epsilon^\mu_{\s\rho\alpha})\mathbb{I}+\frac{i}{4}(\delta^\mu_\alpha\delta^\nu_\beta+\delta^\nu_\alpha\delta^\mu_\beta)\gamma_\rho,\label{S52s}\ee
	and satisfy the relations
	\bea \left[(S^{\mu})^{\alpha\beta\rho\sigma},(S_{\nu})_{\sigma\rho\lambda\gamma}\right]&=&i\epsilon^{\mu}_{\s\nu\delta}(S_{}^{\delta})^{\alpha\beta\rho\sigma},\\
	(S_\rho)^{\mu\nu}_{\s\lambda\sigma}(S^\rho)^{\lambda\sigma}_{\s\alpha\beta}&=&-\alpha(\alpha+1){I}^{\mu\nu}_{\alpha\beta},\quad \alpha=\frac{5}{2},\eea
	with the identity operator  ${I}^{\mu\nu}_{\alpha\beta}$ given by:
	\bea {I}^{\mu\nu}_{\alpha\beta}=\frac{1}{2}(\delta^\mu_\alpha\delta^\nu_\beta+\delta^\nu_\alpha\delta^\mu_\beta)\mathbb{I}-\frac{16}{35}\eta^{\mu\nu}\eta_{\alpha\beta}\mathbb{I}-\frac{4}{35}\left[(\delta^\nu_\beta\gamma^\mu+\delta^\mu_\beta\gamma^\nu)\gamma_\alpha+(\delta^\nu_\alpha\gamma^\mu+\delta^\mu_\alpha\gamma^\nu)\gamma_\beta\right].\eea
	The self-dual description for massive spin-5/2 in $D=2+1$ dimensions was introduced in the work by  Aragone and Stephany in 1984 \cite{aragone spin-5/2}. It is possible to show that, after satisfying the Fierz-Pauli constraints, the equations of motion of these models can be written in terms of (\ref{S52s}).
	
\subsection{helicity - $\alpha=3$}

From (\ref{SB3}), the explicit expression for the symmetric spin generator for the rank $s=3$ case is given by:
	\bea  
(S_\rho)^{\mu\nu\gamma}_{\s\alpha\beta\lambda}&=&\frac{i}{3}(\epsilon_{\alpha\rho}^{\s\s\mu}I^{\nu\gamma}_{\beta\lambda}+\epsilon_{\beta\rho}^{\s\s\mu}I^{\nu\gamma}_{\alpha\lambda}+\epsilon_{\lambda\rho}^{\s\s\mu}I^{\nu\gamma}_{\beta\alpha}\nonumber\\
	&+&\epsilon_{\alpha\rho}^{\s\s\nu}I^{\mu\gamma}_{\beta\lambda}+\epsilon_{\beta\rho}^{\s\s\nu}I^{\mu\gamma}_{\alpha\lambda}+\epsilon_{\lambda\rho}^{\s\s\nu}I^{\mu\gamma}_{\beta\alpha} \nonumber\\
	&+&\epsilon_{\alpha\rho}^{\s\s\gamma}I^{\nu\mu}_{\beta\lambda}+\epsilon_{\beta\rho}^{\s\s\nu}I^{\mu\gamma}_{\alpha\lambda}+\epsilon_{\lambda\rho}^{\s\s\nu}I^{\mu\gamma}_{\beta\alpha}), \label{S de spin-3}    \eea
	where
	\bea {{I}}^{\sigma\phi\omega}_{\mu\nu\rho}&=&-\frac{1}{3}\left(\delta^{\sigma}_{\mu}I^{\phi\omega}_{\nu\rho}+\delta^{\phi}_{\mu}I^{\sigma\omega}_{\nu\rho}+\delta^{\omega}_{\mu}I^{\phi\sigma}_{\nu\rho}\right)\nn\\
	&+&\frac{1}{18}\left[\delta_{\mu\nu}(\delta^{\phi}_{\rho}\delta^{\omega\sigma}+\delta^{\sigma\phi}\delta^{\omega}_{\rho}+\delta^{\sigma}_{\rho}\delta^{\omega\rho})+ (\rho\leftrightarrow\nu) + (\rho\leftrightarrow\mu) \right]. \eea
We can check that the spin generator $S_{\rho}$ in (\ref{S de spin-3}) obeys the following relations: 
	\bea [(S^{\mu})^{\alpha\beta\gamma}_{\s\gamma\lambda\delta},(S_{\nu})^{\gamma\lambda\delta}_{\s\rho\nu\sigma}]&=&i\epsilon^{\mu}_{\s\nu\delta}(S^{\delta})^{\alpha\beta\gamma}_{\s\rho\nu\sigma}\label{alg3}\\
	(S_{}^{\mu})^{\alpha\beta\gamma}_{\s\gamma\lambda\delta}(S_{\mu})^{\gamma\lambda\delta}_{\s\rho\nu\sigma}&=&-\alpha(\alpha+1)I^{\alpha\beta\gamma}_{\rho\nu\sigma} \quad , \quad \alpha=3.\label{alpha3}\eea
	The first-order self-dual model, which describes a massive particle with helicity $+3$ or $-3$ in $D=2+1$ dimensions, was introduced by Aragone and Khoudeir in 1993 in \cite{aragone spin-3 e spin-4}. In \cite{denis e elias nge spin-3}, the authors use the expression (\ref{S de spin-3}) in order to write down the equations of motion coming from that model.

\subsection{helicity - $\alpha=4$}

	From (\ref{SB3}), we can obtain the explicit expression for the symmetric spin generator for the rank $s=4$ case, which is given by
	\bea 
	(S_\rho)^{\mu\nu\alpha\beta}_{\s\lambda\sigma\gamma\varphi}&=&\frac{i}{4}\left[\epsilon^\mu_{\s\rho\lambda}I^{\nu\alpha\beta}_{\sigma\gamma\varphi}+\epsilon^\mu_{\s\rho\sigma}I^{\nu\alpha\beta}_{\lambda\gamma\varphi}+\epsilon^\mu_{\s\rho\gamma}I^{\nu\alpha\beta}_{\sigma\lambda\varphi}+\epsilon^\mu_{\s\rho\varphi}I^{\nu\alpha\beta}_{\sigma\gamma\lambda}\right.\nonumber\\
	&+&\epsilon^\nu_{\s\rho\sigma}I^{\mu\alpha\beta}_{\lambda\gamma\varphi}+\epsilon^\nu_{\s\rho\lambda}I^{\mu\alpha\beta}_{\sigma\gamma\varphi}+\epsilon^\nu_{\s\rho\gamma}I^{\mu\alpha\beta}_{\lambda\sigma\varphi}+\epsilon^\nu_{\s\rho\varphi}I^{\mu\alpha\beta}_{\lambda\gamma\sigma}\nonumber\\
	&+&\epsilon^\alpha_{\s\rho\gamma}I^{\nu\mu\beta}_{\sigma\lambda\varphi}+\epsilon^\alpha_{\s\rho\sigma}I^{\nu\mu\beta}_{\gamma\lambda\varphi}+\epsilon^\alpha_{\s\rho\lambda}I^{\nu\mu\beta}_{\sigma\gamma\varphi}+\epsilon^\alpha_{\s\rho\varphi}I^{\nu\mu\beta}_{\sigma\lambda\gamma}\nonumber\\
	&+& \left. \epsilon^\beta_{\s\rho\varphi}I^{\nu\alpha\mu}_{\sigma\gamma\lambda}+\epsilon^\beta_{\s\rho\sigma}I^{\nu\alpha\mu}_{\varphi\gamma\lambda}+\epsilon^\beta_{\s\rho\gamma}I^{\nu\alpha\mu}_{\sigma\varphi\lambda}+\epsilon^\beta_{\s\rho\lambda}I^{\nu\alpha\mu}_{\sigma\gamma\varphi}\right],\eea
	where
	\bea I^{\mu\nu\alpha\beta}_{\lambda\sigma\gamma\varphi}&=&\frac{1}{4}\Big(\delta^\mu_\alpha I^{\nu\alpha\beta}_{\sigma\gamma\varphi}+\delta^\nu_\alpha I^{\mu\alpha\beta}_{\sigma\gamma\varphi}+\delta^\alpha_\alpha I^{\nu\mu\beta}_{\sigma\gamma\varphi}+\delta^\beta_\alpha I^{\nu\alpha\mu}_{\sigma\gamma\varphi}\Big)\nonumber\\
	&-&\frac{1}{60}\left[\eta_{\lambda\sigma}(\delta^{\mu\nu}I^{\alpha\beta}_{\gamma\varphi}+\delta^{\mu\alpha}I^{\nu\beta}_{\gamma\varphi}+\delta^{\mu\beta}I^{\alpha\nu}_{\gamma\varphi}+\delta^{\nu\alpha}I^{\mu\beta}_{\gamma\varphi}+\delta^{\nu\beta}I^{\mu\alpha}_{\gamma\varphi}+\delta^{\alpha\beta}I^{\mu\nu}_{\gamma\varphi})\right.\nonumber\\
	&+& \left. (\sigma\leftrightarrow\gamma)+(\sigma\leftrightarrow\varphi)+(\lambda\leftrightarrow\gamma)+(\lambda\leftrightarrow\varphi)+(\lambda\sigma\leftrightarrow\gamma\varphi)\right].\eea
	With such expressions at hand, we are able to verify that: 
	\bea
	\left[(S^{\mu})^{\alpha\beta\gamma\phi}_{\s\gamma\lambda\delta\eta},(S_{\nu})^{\gamma\lambda\delta\eta}_{\s\rho\nu\sigma\xi}\right]&=&i\epsilon^{\mu}_{\s\nu\delta}(S^{\delta})^{\alpha\beta\gamma\phi}_{\s\rho\nu\sigma\xi}\label{ss4}\\
	(S_\rho)^{\mu\nu\alpha\beta}_{\s\eta\omega\delta\xi}(S^\rho)^{\eta\omega\delta\xi}_{\s\lambda\sigma\gamma\varphi}&=&-\alpha(\alpha+1)I^{\mu\nu\alpha\beta}_{\lambda\sigma\gamma\varphi}, \quad \alpha =4.\eea
	The first-order self-dual model describing massive particles of helicity $+4$ or $-4$ in $D=2+1$ dimensions was also introduced in \cite{aragone spin-3 e spin-4}. It is possible to write the equations of motion coming from this model in terms of the symmetric generators that we have explicitly deduced here.

\section{Rank-s algebra} \label{app:B}

	Here, we demonstrate the spin algebra for arbitrary $s$. From (\ref{SB3}), we have:
	\bea\left[S_\alpha,S_\beta\right]^{\mu_1...\mu_s}_{\s\nu_1...\nu_s}&=&(S_\alpha)^{\mu_1...\mu_s}_{\s\rho_1...\rho_s}(S_\beta)^{\rho_1...\rho_s}_{\s\nu_1...\nu_s}-(S_\beta)^{\mu_1...\mu_s}_{\s\rho_1...\rho_s}(S_\alpha)^{\rho_1...\rho_s}_{\s\nu_1...\nu_s}\nonumber\\
	&=&i^2\sum\limits^s_{k=1}\prod\limits_{i=1\atop i\neq k}^s\epsilon^{\mu_k}_{\s\alpha\rho_k}\delta^{\mu_i}_{\rho_i}\sum\limits^s_{l=1}\prod\limits_{j=1\atop j\neq l}^s\epsilon^{\rho_l}_{\s\beta\nu_l}\delta^{\rho_j}_{\nu_j}-i^2\sum\limits^s_{k=1}\prod\limits_{i=1\atop i\neq k}^s\epsilon^{\mu_k}_{\s\beta\rho_k}\delta^{\mu_i}_{\rho_i}\sum\limits^s_{l=1}\prod\limits_{j=1\atop j\neq l}^s\epsilon^{\rho_l}_{\s\alpha\nu_l}\delta^{\rho_j}_{\nu_j}. \label{ss-ss}
	\eea
	To analyze this expression, we can consider two scenarios: when $k\neq l$ and when $k=l$. Let's begin with the first case ($k\neq l$). In this scenario, $k$ takes the value of $j$ and $l$ takes the value of $i$, resulting in contractions between Levi-Civita symbols and Kronecker deltas. Therefore, the right hand side of (\ref{ss-ss}) can be expressed as follows:
\be
	\sum\limits^s_{k,l=1\atop k\neq l}\prod\limits^s_{i=1\atop i\neq k,l}\delta^{\mu_i}_{\nu_i}\epsilon^{\mu_k}_{\s\alpha\nu_k}\epsilon^{\mu_l}_{\s\beta\nu_l}-\sum\limits^s_{k,l=1\atop k\neq l}\prod\limits^s_{i=1\atop i\neq k,l}\delta^{\mu_i}_{\nu_i}\epsilon^{\mu_k}_{\s\beta\nu_k}\epsilon^{\mu_l}_{\s\alpha\nu_l}.
 \ee
	Notice that $l$ and $k$ are dummy indices and can be interchanged with another index. We substitute $l$ with $k$ and $k$ with $l$:
\be \sum\limits^s_{k,l=1\atop k\neq l}\prod\limits^s_{i=1\atop i\neq k,l}\delta^{\mu_i}_{\nu_i}\epsilon^{\mu_k}_{\s\alpha\nu_k}\epsilon^{\mu_l}_{\s\beta\nu_l}-\sum\limits^s_{k,l=1\atop k\neq l}\prod\limits^s_{i=1\atop i\neq k,l}\delta^{\mu_i}_{\nu_i}\epsilon^{\mu_l}_{\s\beta\nu_l}\epsilon^{\mu_k}_{\s\alpha\nu_k}=0.\ee
	Now, we can analyze the second case ($k=l$). In this scenario, the sums in (\ref{ss-ss}) can be reduced to just one sum. Additionally, the Kronecker deltas are always contracted, which simplifies the product of symbols to
	\bea \sum^s_{k=1}\prod^s_{i=1\atop i\neq k}\delta^{\mu_i}_{\nu_i}\epsilon^{\mu_k}_{\s\alpha\rho_k}\epsilon^{\rho_k}_{\s\beta\nu_k}-\sum^s_{k=1}\prod^s_{i=1\atop i\neq k}\delta^{\mu_i}_{\nu_i}\epsilon^{\mu_k}_{\s\beta\rho_k}\epsilon^{\rho_k}_{\s\alpha\nu_k} &=& \sum^s_{k=1}\prod^s_{i=1\atop i\neq k}\delta^{\mu_i}_{\nu_i}\left(-\delta^{\mu_k}_\beta\eta_{\alpha\nu_k}+\delta^{\mu_k}_{\nu_k}\eta_{\alpha\beta}+\delta^{\mu_k}_\alpha\eta_{\beta\nu_k}-\delta^{\mu_k}_{\nu_k}\eta_{\alpha\beta}\right) \nonumber \\
 &=& \sum^s_{k=1}\prod^s_{i=1\atop i\neq k}\delta^{\mu_i}_{\nu_i}\left(-\delta^{\mu_k}_\beta\eta_{\alpha\nu_k}+\delta^{\mu_k}_\alpha\eta_{\beta\nu_k}\right).
 \eea

	\no Notice that the term $-\delta^{\mu_k}_\beta\eta_{\alpha\nu_k}+\delta^{\mu_k}_\alpha\eta_{\beta\nu_k}$ can be rewritten as $\epsilon^{\mu_k}_{\rho\nu_k}\epsilon_{\alpha\beta}^{\s\s\rho}$, so we have
	\be\epsilon_{\alpha\beta}^{\s\s\rho}\sum^s_{k=1}\prod^s_{i=1\atop i\neq k}\delta^{\mu_i}_{\nu_i}\epsilon\epsilon^{\mu_k}_{\rho\nu_k}=-i\epsilon_{\alpha\beta}^{\s\s\rho}(S_\rho)^{\mu_1...\mu_s}_{\s\nu_1...\nu_s}.\ee
	Finally, by replacing this result in (\ref{ss-ss}), we have
	\be[S_\alpha,S_\beta]^{\mu_1...\mu_s}_{\s\nu_1...\nu_s}=-i\epsilon_{\alpha\beta}^{\s\s\rho}(S_\rho)^{\mu_1...\mu_s}_{\s\nu_1...\nu_s},\ee
	completing the demonstration.

\end{document}